\documentclass[12pt]{article}
\pdfoutput=1

\usepackage{color}
\usepackage{graphicx}
\usepackage{amsmath}
\usepackage{slashed}
\usepackage{amssymb}
\usepackage{epstopdf}

\input xy
\xyoption{all}
\xyoption{web}

\usepackage[
      colorlinks=true,
      linkcolor=blue,
      urlcolor=blue,
      filecolor=black,
      citecolor=red,
      pdfstartview=FitV,
      pdftitle={},
        pdfauthor={Michael Gutperle, Yi Li},
        pdfsubject={},
        pdfkeywords={},
        pdfpagemode={},
        bookmarksopen=true
      ]{hyperref}

\usepackage{color}
\usepackage{graphicx}

\usepackage{sectsty}
\sectionfont{\large}

\renewcommand{\thefootnote}{\fnsymbol{footnote}}
\renewcommand{\thanks}[1]{\footnote{#1}}
\newcommand{\starttext}{
\setcounter{footnote}{0}
\renewcommand{\thefootnote}{\arabic{footnote}}}

\newcommand{\bea}{\begin{eqnarray}}
\newcommand{\eea}{\end{eqnarray}}
\newcommand{\be}{\begin{equation}}
\newcommand{\ee}{\end{equation}}

 \newcommand{\reals}{\mathbb{R}}

\newcommand{\mc}{\mathcal }

\def\half{ {1\over 2}}

\DeclareMathOperator{\tr}{tr}
\usepackage{ dsfont } 

\long\def\symbolfootnote[#1]#2{\begingroup%
\def\thefootnote{\fnsymbol{footnote}}\footnote[#1]{#2}\endgroup}

\allowdisplaybreaks

\begin{document}
\setlength{\baselineskip}{18pt}

\starttext
\setcounter{footnote}{0}

%
\bigskip

\begin{center}

{\Large \bf  Higher Spin Lifshitz Theories and the KdV-Hierarchy}

\vskip 0.4in

{\large Matteo Beccaria$^a$, Michael Gutperle$^b$, Yi Li$^b$ and Guido Macorini$^a$  }

\vskip .2in

{\it  ${}^a$Dipartimento di Matematica e Fisica Ennio De Giorgi,}\\
{\it  Universit\`a del Salento \& INFN, Via Arnesano, 73100 Lecce, Italy}

\vskip .1in

{ \it   ${}^b$Department of Physics and Astronomy }\\
{\it University of California, Los Angeles, CA 90095, USA}\\[0.5cm]

\href{mailto:matteo.beccaria@le.infn.it}{\texttt{matteo.beccaria@le.infn.it}}\texttt{, }\href{mailto:gutperle@physics.ucla.edu}{\texttt{gutperle@physics.ucla.edu}}\\
\href{mailto:yli@physics.ucla.edu}{\texttt{yli@physics.ucla.edu}}\texttt{, }\href{mailto: macorini@nbi.ku.dk}{\texttt{ macorini@nbi.ku.dk}}

\bigskip

\bigskip

\end{center}

\begin{abstract}

\setlength{\baselineskip}{18pt}

In this paper three dimensional higher spin theories in the Chern-Simons formulation with gauge algebra $sl(N,\reals)$ are investigated which have Lifshitz symmetry with  scaling exponent $z$.  We show that an explicit map exists for all $z$ and $N$ relating  the Lifshitz Chern-Simons theory to the $(n,m)$ element of the KdV hierarchy. Furthermore we show that the map and hence the conserved charges are independent of $z$. We derive these result from the Drinfeld-Sokolov formalism of integrable systems.

\end{abstract}

\setcounter{equation}{0}
\setcounter{footnote}{0}

%
%
%
\newpage
\tableofcontents
\newpage

\allowdisplaybreaks

\section{Introduction}
\setcounter{equation}{0}
\label{sec1}

Higher spin theories  in $d$ space time dimensions were constructed over the last 20 years by Vasiliev and collaborators. For some  reviews see e.g.   \cite{Vasiliev:2000rn,Vasiliev:2004qz,Bekaert:2005vh,Didenko:2014dwa}.  These theories provide new ways to explore the AdS/CFT correspondence \cite{Klebanov:2002ja,Giombi:2009wh}. The present paper only deals with higher spin theory in three dimensions in the Chern-Simons (CS) formulation   \cite{Blencowe:1988gj,Bergshoeff:1989ns,Henneaux:2010xg,Campoleoni:2010zq}.
Gaberdiel and Gopakumar \cite{Gaberdiel:2010pz,Gaberdiel:2012uj} proposed a duality linking dimensional higher spin theories  in three dimensional  anti de-Sitter space to  two dimensional $W_N$ minimal model CFTs.

In the last couple of years solutions of three dimensional higher spin gravity which are  not asymptotically AdS have been investigated  in the literature \cite{Gary:2012ms,Afshar:2012nk,Gonzalez:2013oaa,Gary:2014mca,Gutperle:2013oxa}. In particular  asymptotically Lobachevsky, Schr\" odinger, warped AdS and Lifshitz spacetimes have been found. Field theories which exhibit with Lifshitz scaling, i.e.  anisotropic scaling symmetries  of space and time dimensions, are important condensed matter theories near   quantum critical points (see e.g. \cite{Kachru:2008yh}). 

The goal of the present paper is the generalize the results  \cite{Gutperle:2014aja} where a map  of the  Lifshitz Chern-Simons theories with gauge group $sl(N,\reals)$ and scaling exponent $z$ to  the integrable KdV hierarchy was discovered for particular values of $N,z$, namely $N=3,z=2$ as well as $N=4,z=3$.

The structure and the main results of the paper are as follows: In section \ref{sec2} we review some of the background material  and results from \cite{Gutperle:2014aja} for the convenience of the reader. 

In section \ref{sec3}, a detailed analysis of the case of scaling exponent  $z=2$ for generic $N$ is presented. In addition solutions for scaling exponent  $z>2$ and values $N$ up to $N=8$ are found.
These results give very  strong evidence for the conjecture of  \cite{Gutperle:2014aja}, that there always exists a map which relates the $sl(N,\reals)$ $z$ Lifshitz theory to the $m=z,n=N$ member of the KdV hierarchy.

Furthermore  the case by case study reveals also an unexpected universality: First, the form of the map from the Chern-Simons variables  to the KdV variables is independent of $z$ and second, the  form of
 the conserved charges which is determined for $z = 2$ are conserved for all $z$ (and $N$).
 
 In section \ref{sec4}, we use the formalism of matrix  valued pseudo differential operators  constructed by Drinfeld and Sokolov in their seminal paper  \cite{Drinfeld:1984qv} to proof the relation of the CS Lifshitz and KdV and the universality of the map and the conserved charges for all values of $z$ and $N$.
 
 We discuss some directions for future research in section \ref{sec5}.

In Appendix \ref{appa} we  present our conventions for the gauge algebras. In Appendix \ref{appb} details of some of the proof of statements in the paper of Drinfeld and Sokolov  \cite{Drinfeld:1984qv}  are reviewed to make our paper self-contained. 
Some of the results used in section \ref{sec3} are presented in Appendix \ref{appkdv} where we report the 
$z$-independent map between the CS and KdV variables, as well as the explicit KdV and CS equations of motions
for various pairs $N,z$.

\medskip

\section{Review of higher spin Lifshitz theories}
\setcounter{equation}{0}
\label{sec2}
In this section we will   review the Chern-Simons (CS) formulation of higher spin gravity in three dimensions based on the $sl(N,\reals)$ or $hs(\lambda)$ gauge algebra. More details can be found, for example, in 
\cite{Gaberdiel:2012uj,Ammon:2012wc}. In addition, we review  some the results obtained in previous papers of some of the authors  on the formulation of theories with Lifshitz scaling in higher spin gravity theories \cite{Gutperle:2013oxa} and the relation of these theories to the KdV hierarchy \cite{Gutperle:2014aja}.

\subsection{Chern-Simons formulation of higher spin gravity}
The action for the Chern-Simons formulation of higher spin gravity is given by two copies of Chern-Simons at level $k$ and $-k$ respectively

\be\label{chernsimonsa}
S=S_{CS}[A]-S_{CS}[\bar A]
\ee
where the Chern-Simons action is given by the following expression
\be
S_{CS}[A]= {k\over 4\pi}  \int {\rm tr}\Big( A\wedge dA+{2\over 3} A\wedge A\wedge
A\Big).
\ee
The equations of motion following from the Chern-Simons action are the  flatness conditions on the connections $A,\bar A$
\be\label{flata}
F=dA+ A\wedge A=0, \quad \quad \bar F=d\bar A+ \bar A\wedge \bar A=0.
\ee
The gauge connections can be related to generalizations of the vielbein and the spin connection, which take values in the gauge algebra
\be
e_\mu={l\over 2} (A_\mu-\bar A_\mu), \quad \omega_\mu = {1\over 2} (A_\mu+\bar A_\mu).
\ee
The metric and the higher spin fields can be obtained from the vielbein. For example for the $sl(3,\reals)$ case one gets \cite{Campoleoni:2010zq}
\be\label{metviel}
g_{\mu\nu}={1\over 2} \tr ( e_\mu e_\nu), \quad \phi_{\mu\nu\rho}= {1\over 6} \tr(e_{(\mu} e_\nu e_{\rho)}).
\ee
Generalizations of these expressions for $sl(N,\reals)$ were obtained in   \cite{Campoleoni:2012hp,Campoleoni:2014tfa}. In the following we will only need the expression for the metric which is given by (\ref{metviel}). An important ingredient to construct spacetimes with a given asymptotic behavior and their symmetry,  is the   radial gauge. 
We denote a radial coordinate
$\rho$, where the holographic boundary will be located  at $\rho\to \infty$. The coordinates 
$t$ and  $x$ have  the topology of $\reals\times S^1$ or $\reals\times \reals$.   The dependence of the connections $A,\bar A$ on the radial coordinate $\rho$ is given by a gauge transformation on   $\rho$ independent connections $a,\bar a$ 
\be\label{bigadef}
A_\mu = b^{-1} a_\mu  \,b + b^{-1}\partial_\mu b, \qquad \bar A_\mu = b \,\bar
a_\mu b^{-1} + b
\,\partial_\mu  (b^{-1}).
\ee
Where  $b= \exp( \rho L_0)$ and $L_0$ is given by  a  Cartan generator of a $sl(2,\reals)$ sub-algebra of $sl(N,\reals)$. For  $hs(\lambda)$ one chooses the  generator $V^2_0$ instead.
The nonzero components $a_t,a_x$  (and $\bar a_t,\bar a_x$)  obey  the $\rho$ independent flatness condition
\be\label{flatb}
\partial_t a_x -\partial_x a_t + [a_t,a_x]=0, \quad \partial_t \bar{a}_x -\partial_x \bar{a}_t + [\bar{a}_t,\bar{a}_x]=0.
\ee
It is easy to see that connections  satisfying (\ref{flatb}) also satisfy (\ref{flata}).

\subsection{Lifshitz scaling in field theories}
Scaling symmetries are ubiquitous in two dimensional quantum field theories  and generated by the transformation
\be\label{lifscale}
  t\to \lambda^z t, \qquad x\to \lambda x.
\ee
The case $z=1$ corresponds to isotropic scaling and leads to conformally invariant theories. For $z\neq 1$ the scaling is anisotropic and called Lifshitz scaling with exponent $z$. While such an anisotropic scaling breaks Lorentz symmetry it nevertheless appears in some condensed matter systems (see e.g. 
\cite{Kachru:2008yh}). 
The algebra of Lifshitz symmetries is generated by the generator of dilations  $D$ together with  the generator of time  translations $H$ and
spatial
translations $P$. Together they satisfy the following algebra
\bea
  {[P,H]}&=&0 \nonumber\\
  {[D,H]}&=&z H \nonumber\\
  {[D,P]}&=&P .\label{lifalg}
\eea
 The  stress-energy tensor  for field theories in 1+1
dimensions with
Lifshitz scaling is not necessarily symmetric and contains four components:  the energy density
${\cal E}$, the
energy flux ${\cal E}^x$, the momentum density ${\cal P}_x$ and the stress energy
$\Pi_x^{\;x}$.
They satisfy the following conservation equations 
\cite{Ross:2011gu}
\bea \label{emcoma}
  \partial_t {\cal E}+ \partial_x {\cal E}^x=0,  \qquad
  \partial_t {\cal P}_x+ \partial_x {\Pi}_x^{\; x}=0.
\eea
For theories with the Lifshitz scaling exponent $z$ there exists a modified trace condition
\be\label{emcomb}
  z {\cal E}+ {\Pi}_x^{\; x}=0.
\ee

\subsection{Lifshitz spacetimes in higher spin gravity}
A holographic realization of the Lifshitz scaling symmetry in three dimensions can be constructed using the following metric
\be\label{lifmetb}
 ds^2=d\rho^2-e^{2z\rho}dt^2+e^{2\rho}dx^2.
\ee
A shift of the radial coordinate  $\rho \to \rho + \ln \lambda$ induces a Lifshitz scaling transformation on the space time coordinates $t,x$ with scaling exponent $z$  (\ref{lifscale}). Such a metric is in general not a solution of pure Einstein gravity with a negative cosmological constant and additional matter has to be added to support the solution (see e.g. \cite{Kachru:2008yh}). In higher spin gravity Lifshitz metrics can be obtained from connections.
\be\label{lifshconc}
a_{Lif}= V_{\;z}^{z+1} dt + V_1^2 dx, \quad  \bar a_{Lif}= V^{z+1}_{-z} dt + V^2_{\;-1} dx.
\ee 
Since $z$ in general is an integer these constructions produce Lifshitz theories with integer scaling 
exponent. 
Note that the barred connection in  (\ref{lifshconc}) can be related to the unbarred sector by a 
conjugation operation $\bar A= A^c$, where  the conjugation is acting on the gauge algebra generator by $
(V^s_m)^c=(-1)^{s+m+1}V^s_{-m}$. Though in general $A$ and $\bar{A}$ maybe unrelated, $A^c$ solves the flatness condition in the barred sector if $A$ solves it in the unbarred sector so we always take $\bar{A}$ to be $A^c$ as in \cite{Gutperle:2014aja}. By this choice we can get the Lifshitz metric from (\ref{metviel}).

\subsection{Asymptotic Lifshitz connections}\label{sec24}

In holographic theories one considers spacetimes which are not exactly AdS, but approach AdS asymptotically. This enlarges the space of possible solutions including  for example black holes. For Lifshitz spacetimes a similar notion exists. In the Chern-Simons formulation we call a connection asymptotically Lifshitz if  the leading term of the connection is the Lifshitz connection which can be obtained from (\ref{lifshconc}).
In  \cite{Gutperle:2014aja}  
we presented a general procedure to construct time dependent asymptotically Lifshitz connections. The starting point is to choose a "lowest weight gauge" for the connection $a_x$
\be\label{axconn}
a_x = V^2_1 + \sum_{i=2}^\infty \alpha_i V^i_{-i+1} 
\ee
where the $\alpha_i$ depend on $x,t$. An ansatz for the time component of the connection for a asymptotically Lifshitz connection with exponent $z$ is given by 
\be\label{atconn}
a_t = (*a_x)^z |_{traceless} + \Delta a_t.
\ee
In  \cite{Gutperle:2014aja} it was shown that the flatness conditions  (\ref{flatb}) together with $\Delta a_t$ 
can be solved recursively. While the general procedure was developed for $hs(\lambda)$, explicit expressions for two cases, namely $sl(3,\reals), z=2$ and $sl(4,\reals), z=3$ were given in that paper.
In these specific examples it was found that there is some gauge freedom left in the $\Delta a_t$. By appropriately fixing $a_t$ we obtained the equation of motion for $\alpha_i$'s which can be mapped to KdV hierarchy. Another useful property of the CS construction is the fact that one can assign scaling dimensions to the fields $a_i$. The scaling behavior is determined by demanding that under Lifshitz scaling of the coordinates $x\to \lambda x, t\to \lambda^z t$, the connection $A$ is invariant. A field of scaling dimension $l$ will be rescaled by a factor $\lambda^{-l}$.  It was shown in  \cite{Gutperle:2014aja}, that one can assign the following scaling dimensions to the basic fields and operators
\be
[\alpha_n]=n, \quad [\partial_x] =1, \quad [\partial_t]=z.
\ee

\subsection{ Integrability and map to KdV hierarchy} 
Here we briefly describe the formulation of the KdV hierarchy using pseudo differential operators.
Elements of KdV hierarchy are labeled by two integers  $n$ and $m$. A differential operator $L$ can be defined
\be
L= \partial^n+ u_2 \partial^{n-2} +\cdots +u_{n-1} \partial + u_n.
\ee
Here $\partial= {\partial \over \partial x}$ and $u_i=u_i(x,t)$. The formalism of pseudo differential operators  (PDO) introduces negative powers $\partial^{-k}$ of differentiation while preserving the standard rules of differentiation such as the Leibniz rule (see   \cite{dickey,battle92} for reviews). This formalism makes it possible  to define  fractional powers of $L$, in particular $L^{1/n}$.
\be
L^{1/ n} = \partial + {1\over n} u \partial^{-1}+ o(\partial^{-2}).
\ee
For another integer $m$ one defines
\be
P_m = \Big(L^{m/n}\Big)_+
\ee
where the subscript $()_+$ denotes the non-negative part of the pseudo differential operator, which has terms with $\partial^k, k\geq 0$. An integrable system is constructed due to the fact that $P,L$ form a Lax pair, i.e. the evolution equation
\be\label{laxpair}
{\partial\over \partial t}L=[P_m ,L] 
\ee 
gives a system of partial differential equations  for $u_i(x,t)$ which is integrable. In  \cite{Gutperle:2014aja} it was found that for the concrete example $sl(3,\reals), z=2$ and $sl(4,\reals),z=3$ it was possible for  a specific gauge choice  for $a_t$ (called KdV gauge) to map the flatness conditions for the asymptotically Lifshitz connection to the evolution equation (\ref{laxpair}) of an element of KdV hierarchy. Furthermore, it was conjectured that this hold in general with the identification of Chern Simons parameters $N,z$
with the KdV parameters $m,n$ given by
\be
m=z, \quad n=N.
\ee

\section{Explicit Chern-Simons  to KdV maps}
\setcounter{equation}{0}
\label{sec3}

In this section, we illustrate the specific form of the CS-KdV map in various explicit examples. 

\subsection{$z=2$}

A particularly simple case is when the exponent $z$ takes its minimal non-trivial value $z=2$.
We can write the equations of motion for the CS fields $\alpha_{i}$ 
and for the KdV fields $u_{i}$
in closed form for generic $N$. For the $\alpha_{i}$ fields, we have 
\footnote{The second term on the right hand side is zero for $n=2$.}
\be
	\label{alpha-eom}
	\begin{split}
	\dot \alpha_{n} &= \frac{n\,(n^{2}-N^{2})}{2n+1}\,\alpha_{n+1}'+\frac{1}{(n-1)(2n-3)}\,\alpha_{n-1}'''
	+\sum_{m=2}^{n-1}
	\frac{2\,(2n-m)}{2\,(n-m)+1}
	\,\alpha_{n-m+1}'\,\alpha_{m}.
	\end{split}
\ee
For the $u_{i}$ fields of the KdV $(2, N)$ hierarchy, we have 
\be
	\dot u_{i} = u_{i}'' + 2 u_{i+1}' - \frac{2}{N} \, \binom{N}{i}\, u_2^{(i)} - \sum_{j=2}^{i-1}
	{\frac{2}{N} \binom{N-j}{i-j}\, \,u_j \, u_2^{(i-j)} }.
	\label{kdvz2}
\ee
Assuming an Ansatz for the map consistent with the scaling ({\em i.e.} writing the generic 
$u_i$ as a the most general linear combination of $\alpha$'s independent monomials with the correct dimension),
 the matching between the two set of equations can be solved recursively term by term. The 
 explicit expression of the map for the first seven fields turns
 out  to be:
\be\label{map1}
    \begin{split}
    u_{2} &= N(N^{2}-1)\,\frac{\alpha _2}{6}, 
    \end{split}
\ee
\be\label{map2}
\begin{split}
u_{3} &= N(N^{2}-1)(N-2)\,\bigg(
\frac{1}{30} \alpha _3 (-N -2)+\frac{\alpha _2'}{12}
\bigg), 
\end{split}
\ee
\be\label{map3}
\begin{split}
u_{4} &= N(N^{2}-1)(N-2)(N-3)\,\bigg(
\frac{1}{60} (-N -2) \alpha _3'+\frac{1}{360} \alpha _2^2 (5 N
   +7)\\
   &+\frac{1}{140} \alpha _4 (N +2) (N +3)+\frac{\alpha _2''}{40}
\bigg), 
\end{split}
\ee
\be\label{map4}
\begin{split}
u_{5} &= N(N^{2}-1)(N-2)(N-3)(N-4)\,\bigg(
\frac{1}{360} \alpha _2 (5 N +7) \alpha _2'+\frac{1}{280} (N +2) (N +3)
   \alpha _4'\\
   &+\frac{1}{210} (-N -2) \alpha _3''-\frac{\alpha _2 \alpha _3 (N
   +2) (7 N +13)}{1260}-\frac{1}{630} \alpha _5 (N +2) (N +3) (N
   +4)+\frac{\alpha _2{}^{(3)}}{180}
\bigg), 
\end{split}
\ee
\be\label{map5}
\begin{split}
u_{6} &= N(N^{2}-1)(N-2)(N-3)(N-4)(N-5)\,\bigg(
\frac{\alpha _2^3 \left(35 N ^2+112 N +93\right)}{45360}\\
&+\frac{\alpha _3^2
   (N +2) \left(7 N ^2+34 N +44\right)}{12600}-\frac{\alpha _2 (N
   +2) (7 N +13) \alpha _3'}{2520}+\frac{(7 N +10) \left(\alpha
   _2'\right){}^2}{2016}\\
   &-\frac{\alpha _3 (N +2) (7 N +13) \alpha
   _2'}{2520}-\frac{(N +2) (N +3) (N +4) \alpha _5'}{1260}+\frac{\alpha
   _2 (21 N +29) \alpha _2''}{5040}\\
   &+\frac{(N +2) (N +3) \alpha
   _4''}{1008}+\frac{\alpha _4 \alpha _2 (N +2) (N +3) (3 N
   +7)}{2520}+\\&\frac{\alpha _6 (N +2) (N +3) (N +4) (N
   +5)}{2772}
   +\frac{\alpha _3{}^{(3)} (-N -2)}{1008}+\frac{\alpha _2{}^{(4)}}{1008}
\bigg), 
\end{split}
\ee
\be\label{map6}
\begin{split}
u_{7} &= N(N^{2}-1)(N-2)(N-3)(N-4)(N-5)(N-6)\,\bigg(
\frac{\alpha _2^2 \left(35 N ^2+112 N +93\right) \alpha
   _2'}{30240}\\
   &+\frac{\alpha _3 (N +2) \left(7 N ^2+34 N +44\right)
   \alpha _3'}{12600}-\frac{\alpha _3 \alpha _2^2 (N +2) \left(35 N ^2+144
   N +157\right)}{75600}\\
   &-\frac{\alpha _3 \alpha _4 (N +2) (N +3)
   \left(11 N ^2+65 N +106\right)}{46200}+\frac{(7 N +10) \alpha _2'
   \alpha _2''}{3360}\\
   &+\frac{\alpha _2 (N +2) (N +3) (3 N +7) \alpha
   _4'}{5040}+\frac{\alpha _4 (N +2) (N +3) (3 N +7) \alpha
   _2'}{5040}-\\&\frac{(N +2) (21 N +40) \alpha _2' \alpha
   _3'}{15120}+\frac{(N +2) (N +3) (N +4) (N +5) \alpha
   _6'}{5544}-\frac{\alpha _2 (N +2) (6 N +11) \alpha _3''}{7560}-
   \\&\frac{\alpha
   _3 (N +2) (21 N +38) \alpha _2''}{25200}-\frac{(N +2) (N +3)
   (N +4) \alpha _5''}{4620}-\\&\frac{\alpha _5 \alpha _2 (N +2) (N +3)
   (N +4) (11 N +31)}{41580}+\frac{\alpha _2{}^{(3)} \alpha _2 (14 N
   +19)}{15120}\\
   &-\frac{\alpha _7 (N +2) (N +3) (N +4) (N +5)
   (N +6)}{12012}+\frac{\alpha _4{}^{(3)} (N +2) (N
   +3)}{5040}\\
   &+\frac{\alpha _3{}^{(4)} (-N -2)}{6048}+\frac{\alpha _2{}^{(5)}}{6720}
\bigg). \\
\end{split}
\ee
As expected, 
these equations truncate for positive integer $N$ and define a differential map between the first $N$ CS and KdV fields.

\subsection{Generic $z>2$ and universality of the map}

To analyse cases with $z>2$, we begin by briefly recalling the algorithmic construction of CS solutions with asymptotic Lifshitz scaling 
presented in \cite{Gutperle:2014aja}. 
The main ingredient are the equations (\ref{axconn}), (\ref{atconn}) for the two components  of the connection.
As discussed in section \ref{sec24}, it is convenient to assign 
the scaling dimension $[\alpha_{n}]=n$ to the fields and $[V^{s}_{m}]=m$ to the generators. This implies that 
all terms in (\ref{axconn}) have the same dimension 1. Let $\Phi(\alpha)$ be the set of monomials built with the $\alpha$-functions and their $\partial_{x}$ derivatives. 
Then we can write the following explicit ansatz for  $a_{t}$
\be
	\label{axconn2}
	a_{t} = (*a_x)^z |_{traceless}+\sum_{n=2}^{\infty}\sum_{m=-n+1}^{z-2}\mc O^{n}_{m}(\alpha)\,
	V^{n}_{m},
\ee
where 
$\mc O^{n}_{m}(\alpha)$ is a linear combination of elements of $\Phi(\alpha)$
with homogeneous dimension $z-m$. The upper bound on $m$ is due to the fact that the minimal dimension is 2, obtained for $\mc O\sim \alpha_{2}$. Solving the flatness condition amounts to solve algebraic equations for the coefficients in the $\mc O$ combinations in (\ref{axconn2}). 
This system
has a triangular structure and can be fully reduced to a finite dimensional one for $\lambda=N$ when $hs(\lambda)$
reduces to $sl(N,\reals)$.

For our purposes, it is important to revisit the  case $z=3,\,N=4$ that has already been discussed 
in \cite{Gutperle:2014aja}. 
Solving the Ansatz (\ref{axconn2}), we obtain the following non-zero polynomials $\mc O^{n}_{m}$:
\be
\begin{split}
\mc O^{2}_{-1} &= \bigg(\frac{41}{5}-k\bigg)\,\alpha_{2}^{2}-\frac{k}{2}\,\alpha_{2}'', \quad
\mc O^{2}_{0} = k\,\alpha_{2}', \quad
\mc O^{2}_{1} = \bigg(\frac{41}{5}-k\bigg)\,\alpha_{2}, \\
\mc O^{3}_{-2} &= \bigg(\frac{41}{5}-k\bigg)\,\alpha_{2}\alpha_{3}-\frac{1}{2}\,\alpha_{3}'', \quad
\mc O^{3}_{-1} = 2\,\alpha_{3}', \\
\mc O^{4}_{-3} &= \bigg(\frac{41}{5}-k\bigg)\,\alpha_{2}\alpha_{4}+\frac{3}{10}(\alpha_{2}')^{2}+\frac{23}{60}
\,\alpha_{2}\alpha_{2}''+\frac{1}{10}\,\alpha_{4}''+\frac{1}{120}\,\alpha_{2}'''',\\
\mc O^{4}_{-2} &= -\frac{9}{5}\,\alpha_{2}\,\alpha_{2}'-\frac{3}{5}\,\alpha_{4}'-\frac{1}{20}\,\alpha_{2}''', \quad
\mc O^{4}_{-1} = \frac{1}{4}\,\alpha_{2}'', \quad
\mc O^{4}_{0} = -\alpha_{2}',
\end{split}
\ee
where $k$ is an undetermined coefficient. The associated equations of motion are 
\be
\label{5.2}
\begin{split}
\dot \alpha_{2} &= -3\,k\,\alpha_{2}\alpha_{2}'-\frac{k}{2}\,\alpha_{2}'''+\frac{54}{5}\,\alpha_{4}', \\
\dot \alpha_{3} &= -3\bigg(k+\frac{9}{5}\bigg)\alpha_{3}\alpha_{2}'-\bigg(k+\frac{34}{5}\bigg) \alpha_{2}\alpha_{3}'
-\frac{1}{2}\alpha_{3}''', \\
\dot \alpha_{4} &= 
-\bigg(k-\frac{14}{5}\bigg)\alpha_{4}'\alpha_{2}
-2\bigg(2k-\frac{7}{5}\bigg)\alpha_{4}\alpha_{2}'
+\frac{24}{5}\alpha_{2}'\alpha_{2}^{2}
+\frac{13}{30}\alpha_{2}'''\alpha_{2}\\
&
-12\alpha_{3}\alpha_{3}'
+\frac{59}{60}\alpha_{2}'\alpha_{2}''
+\frac{1}{10}\alpha_{4}'''
+\frac{1}{120}\alpha_{2}'''''.
\end{split}
\ee
These can be compared with the equations of motions quoted in \cite{Gutperle:2014aja}
\be
\label{5.3}
\begin{split}
\dot \alpha_{2} &= -(\frac{123}{5}-3c)\,\alpha_{2}\alpha_{2}'-(\frac{41}{10}-\frac{c}{2})\,\alpha_{2}'''+\frac{54}{5}\,\alpha_{4}', \\
\dot \alpha_{3} &= -(30-3c)\alpha_{3}\alpha_{2}'-(15-c)\,\alpha_{2}\alpha_{3}'
-\frac{1}{2}\alpha_{3}''', \\
\dot \alpha_{4} &= 
-\bigg(\frac{27}{5}-c\bigg)\alpha_{4}'\alpha_{2}
-(30-4c)\,\alpha_{4}\alpha_{2}'
+\frac{24}{5}\alpha_{2}'\alpha_{2}^{2}
+\frac{13}{30}\alpha_{2}'''\alpha_{2}\\
&
-12\alpha_{3}\alpha_{3}'
+\frac{59}{60}\alpha_{2}'\alpha_{2}''
+\frac{1}{10}\alpha_{4}'''
+\frac{1}{120}\alpha_{2}''''',
\end{split}
\ee
where $c$ is a gauge parameter analogous to $k$. If we set 
\be
\label{5.4}
k = \frac{41}{5}-c,
\ee
then equations (\ref{5.2}) and (\ref{5.3}) match. They must be compared with the KdV equations for the $(4,3)$ case, i.e.
\bea\begin{split}
  \dot u_2 & =  -\frac{3}{4} \,u_2  \,u_2' +3 \,u_4' -\frac{3}{2} \,u_3'' +\frac{1}{4} \,u_2''', \\
  \dot u_3 & =  -\frac{3}{4} \,u_3  \,u_2' -\frac{3}{4} \,u_2  \,u_3' +3 \,u_4'' -2 \,u_3''' +\frac{3}{4} \,u_2^{\text{(4)}},  \\
  \dot u_4 & =  -\frac{3}{4} \,u_3  \,u_3' +\frac{3}{4} \,u_2  \,u_4' +\frac{3}{8} \,u_3  \,u_2'' -\frac{3}{4} \,u_2  \,u_3'' +\frac{3}{8} \,u_2 
   \,u_2''' +\,u_4''' -\frac{3}{4} \,u_3^{\text{(4)}} +\frac{3}{8} \,u_2^{\text{(5)}}.
\end{split}\eea
We can try to relate the CS and KdV equations of motion by postulating a generic CS-KdV map 
consistent with the scaling dimensions. In this case, it reads
\be
\begin{split}
u_{2} &= \xi_{2,1}\,\alpha_{2}, \\
u_{3} &= \xi_{3,1}\,\alpha_{3}+\xi_{3,2}\,\alpha_{2}', \\
u_{4} &= \xi_{4,1}\,\alpha_{2}^{2}+\xi_{4,2}\,\alpha_{4}+\xi_{4,3}\alpha_{3}'+\xi_{4,4}\,\alpha_{2}''.
\end{split}
\ee
Comparing the CS and KdV sides, we get a set of algebraic equations for $k$ and the $\xi$-coefficients
which have the following two non-trivial solutions ($\xi\equiv 0$ is clearly a solution):
\be
\label{5.8}
\begin{split}
& \xi_{2,1}=10,\ \xi_{3,1}=\pm 24,\ \xi_{3,2}=10,\ \xi_{4,1}=9, \\
&\xi_{4,2}=36,\ \xi_{4,3}=\pm 12,\ \xi_{4,4}=3, \ k=\frac{7}{10}.
\end{split}
\ee
The value of $k$ implies $c=15/2$ as in \cite{Gutperle:2014aja}, see (\ref{5.4}). However, the solution quoted in that reference is the one with the plus sign in (\ref{5.8}). \footnote{The 
extra map results from the symmetry of the CS equations of motion under the discrete 
transformation $\alpha_i \rightarrow (-1)^i \alpha_i$.} Taking instead the minus sign, we recover precisely
the KdV map for $N=4$ and $z=2$ as one can easily see just taking equations (\ref{map1}-\ref{map6}) for $N=4$.
This simple remark
suggests that the CS-KdV map is actually {\em universal}, {\em i.e.} independent on the Lifshitz 
exponent $z$. We have systematically explored the map for various $z$ and $N=4,5,6,7,8$. 
The explicit results for the CS-KdV maps and the equations of motions  are collected in  Appendix \ref{appkdv}.
One can check that in all cases there is always one solution to the algebraic 
constraints  such that the CS-KdV map is the same as for $z=2$.

\subsection{Conserved charges}

A further check of universality of the CS-KdV map is provided by the conserved charges. In particular, 
we expect that the charges determined for $z=2$ are conserved for all $z$ (and $N$). The explicit form of the 
conserved charges for $z=2$ can be determined by using the closed form of the equations of motion.
Guided by the results of \cite{Gutperle:2014aja}, we look for densities $\rho_{n}$ 
of the form
\be 
\rho_{n} = \alpha_{n}+\text{other fields of dimension n},
\ee
 such that, using (\ref{alpha-eom}), we get 
\be
\partial_{t}\rho_{n} = \partial_{x} ( \text{local field of dimension n+1}).
\ee
At each $n$, we find by direct inspection, a unique solution 
 up to total 
derivatives of previously determined densities $\rho_{m<n}$. The first expressions are trivial
\be
\rho_{2} = \alpha_{2}, \qquad \rho_{3} = \alpha_{3}.
\ee
The next charges have an explicit $N$ dependence and read
\bea
\rho_{4} &=& \alpha _4-\frac{7 \alpha _2{}^2}{6 \left(N ^2-4\right)}, \notag \\
\rho_{5} &=& \alpha _5-\frac{4 \alpha _2 \alpha _3}{N ^2-9}, \notag \\
\rho_{6} &=& \alpha _6-\frac{11 \left(2 N ^2-11\right) \left(\alpha _2'\right){}^2}{24 \left(N
   ^2-16\right) \left(N ^2-9\right) \left(N ^2-4\right)}+\frac{11 \alpha
   _2{}^3 \left(13 N ^2-61\right)}{36 \left(N ^2-16\right) \left(N
   ^2-9\right) \left(N ^2-4\right)}\notag \\
   &&-\frac{11 \alpha _4 \alpha _2}{2 \left(N
   ^2-16\right)}-\frac{11 \alpha _3{}^2 \left(3 N ^2-20\right)}{10 \left(N
   ^2-16\right) \left(N ^2-9\right)}, \notag \\
\rho_{7} &=& \alpha_{7}+\frac{143 \alpha _3 \alpha _2''}{50 \left(N ^2-25\right) \left(N
   ^2-16\right)}+\frac{572 \alpha _3 \alpha _2{}^2}{25 \left(N ^2-25\right)
   \left(N ^2-16\right)}-\frac{104 \alpha _5 \alpha _2}{15 \left(N
   ^2-25\right)}  \\
   && -\frac{13 \alpha _3 \alpha _4 \left(17 N ^2-173\right)}{25
   \left(N ^4-41 N ^2+400\right)}, \notag \\
\rho_{8} &=& \alpha_{8}+\frac{13 \alpha _4 \left(17 N ^2-227\right) \alpha _2''}{60 \left(N
   ^2-36\right) \left(N ^2-25\right) \left(N ^2-16\right)}+\frac{13 \alpha
   _4 \alpha _2{}^2 \left(161 N ^2-2441\right)}{60 \left(N ^2-36\right)
   \left(N ^2-25\right) \left(N ^2-16\right)}\notag \\
   &&-\frac{25 \alpha _6 \alpha
   _2}{3 \left(N ^2-36\right)}-\frac{13 \left(5 N ^4-93 N
   ^2+388\right) \left(\alpha _3'\right){}^2}{30 \left(N ^2-36\right)
   \left(N ^2-25\right) \left(N ^2-16\right) \left(N
   ^2-9\right)}\notag \\
   &&-\frac{143 \alpha _2{}^2 \left(38 N ^4-605 N ^2+1887\right)
   \alpha _2''}{720 \left(N ^2-36\right) \left(N ^2-25\right) \left(N
   ^2-16\right) \left(N ^2-9\right) \left(N ^2-4\right)}\notag \\
   &&-\frac{143 \left(3
   N ^4-50 N ^2+167\right) \left(\alpha _2''\right){}^2}{720 \left(N
   ^2-36\right) \left(N ^2-25\right) \left(N ^2-16\right) \left(N
   ^2-9\right) \left(N ^2-4\right)}\notag \\
   &&-\frac{143 \alpha _2{}^4 \left(281 N
   ^4-4210 N ^2+12569\right)}{2160 \left(N ^2-36\right) \left(N
   ^2-25\right) \left(N ^2-16\right) \left(N ^2-9\right) \left(N
   ^2-4\right)}\notag \\
   &&+\frac{13 \alpha _3{}^2 \alpha _2 \left(97 N ^4-1989 N
   ^2+9692\right)}{30 \left(N ^2-36\right) \left(N ^2-25\right)
   \left(N ^2-16\right) \left(N ^2-9\right)}-\frac{3 \alpha _4{}^2
   \left(271 N ^4-7315 N ^2+54684\right)}{140 \left(N ^2-36\right)
   \left(N ^2-25\right) \left(N ^2-16\right)}\notag \\
   &&-\frac{4 \alpha _3 \alpha _5
   \left(41 N ^2-596\right)}{15 \left(N ^4-61 N ^2+900\right)}.\notag
\eea
These have been derived using the $z=2$ equations of motion. However, 
since the conserved charges on the KdV side are by definition $z$-independent, we 
expect that these expressions are valid for any $z$ as well. Indeed, 
we checked that the above densities define conserved charges for all the examples we explored,
using  the $\alpha_{i}$ equations of motion collected in Appendix \ref{appkdv}.




\section{Integrability of Lifshitz Chern-Simons theory by Drinfeld-Sokolov formalism}
\setcounter{equation}{0}
\label{sec4}

In the preceding sections we have shown 
that Lifshitz Chern-Simons theory is an integrable system by appropriately choosing $a_t$. Though this fact can be verified by constructing the explicit map between the Lifshitz Chern-Simons theory and the KdV hierarchy, an elegant theoretic approach is desired. To begin with, we rewrite the flatness condition in a Lax form
\be
 \frac{d}{dt} D_x + [a_t,D_x]=0,
\ee
where the covariant derivative $D_x = \partial_x + a_x$ is regarded as a Lie algebra valued pseudo differential operator (abbreviated as PDO from now on). For gauge algebra $sl(N,\mathbb{R})$, we can use the matrix representation and the flatness condition becomes a Lax equation of matrix valued PDO. This fact inspires us to look for a formulation of integrable systems in terms of matrix valued PDOs. This formalism was developed in  the seminal paper by Drinfeld and Sokolov \cite{Drinfeld:1984qv}. One of our main results is that both the  Lifshitz Chern-Simons theory for $sl(N,\mathbb{R})$ and the KdV hierarchy can be deduced from the Drinfeld Sokolov formalism and are related  by making two different gauge choices for the PDOs. Consequently, almost all the questions previously studied about integrability of our Lifshitz Chern-Simons theory for the  gauge algebra $sl(N,\mathbb{R})$, including the map from Lifshitz Chern-Simons theory to KdV, the infinite tower of conserved quantities and the choice of $a_t$ to make Lifshitz Chern-Simons theory integrable, are given clear answers.

The Drinfeld Sokolov formalism starts by defining the PDO valued in $sl(N,\mathbb{R})$
\be
 L=\partial_x + q(x,t) + \Lambda,
\ee
where $q$ is a lower triangular matrix (or non-positive weight element, if we use the terminology in $hs(\lambda)$ and view $sl(N,\mathbb{R})$ as a truncation of it)
and\footnote{ The parameter $\lambda$ was introduced by Drinfeld and Sokolov and should not be confused with the deformation  parameter in the gauge algebra $hs(\lambda)$.}
\be
 \Lambda=V^2_1+\lambda e.
\ee
Let $e_{i,j}$ denote the matrix with a single one in the i'th row and j'th column, and zeros elsewhere. In the matrix representation we use, $V^2_1=\sum_{i=1}^{N-1}e_{i,i+1}$, and $e=e_{N,1}$ is proportional to $V^{N}_{-N+1}$. The Lax equation is defined as
\be
 \frac{d}{dt}L=[P,L],
\ee
where $P$ is some differential polynomial in $q$ that has to be carefully chosen. The left hand side of the Lax equation is independent on $\lambda$ and lower triangular, so we want the commutator on the right hand side to be also independent on $\lambda$ and lower triangular. Suppose $M=\sum_{i=-\infty}^n m_i \lambda^i$ is a matrix that commutes with $L$ where $m_i$'s are matrix coefficients (matrices multiplied to the left of powers in $\Lambda$), then we can set $P=M_+$, the part of $M$ with non-negative powers in $\lambda$. From $[M,L]=0$ it follows $[M_+,L]=-[M_-,L]$. Since the left hand side only contains non-negative powers in $\lambda$ but the right hand side only contains non-positive powers in $\lambda$, they should be both independent on $\lambda$ and $-[M_-,L]=[m_{-1},e]$ is necessarily lower triangular. Now we have $[P,L]=[M_+,L]=[m_0,\partial_x+V^2_1+q]$. We identify $V^2_1+q$ as $a_x$, so we have $L=D_x + \lambda e$. We furthermore identify $-m_0=-\textit{Zero}(P)$ as $a_t$, where symbolically $\textit{Zero}$ means to take the $\lambda^0$ part. Then the Lax equation is reduced to our flatness condition in Chern-Simons Lifshitz theory. It should be noted that the parameter $\lambda$ is used in setting up the PDOs, the actual equations of motion and the conserved charges are all independent on $\lambda$.

An important restriction we want to impose on the Lax equation is that it must preserve gauge equivalence. Furthermore it will be shown that the Lifshitz Chern-Simons theory and the KdV hierarchy are just reduction of Drinfeld Sokolov formalism by special gauge choices. The crucial notion, a gauge transformation, is defined for a PDO as
\be
 L^\prime=S^{-1}LS,
\ee
where $S$ is a $\lambda$-independent lower triangular matrix with ones in the diagonal, or in the higher spin algebra language, $S$ is $V^1_0$ plus negative weight terms. Define $L^\prime=\partial_x+a_x^\prime+\lambda e=\partial_x+V^2_1+q^\prime+\lambda e$, then this PDO gauge transformation induces a transformation of $a_x$ (or $q$)
\bea
 a_x^\prime&=&S^{-1}a_x S+S^{-1}\partial_x S, \nonumber \\
 q^\prime&=&S^{-1} V^2_1 S - V^2_1 + S^{-1} \partial_x S,
\eea
where we used the fact that $e$ commutes with $S$ in the calculation. By the explicit construction specified later $P$ is a differential polynomial in $q$ and so is the commutator $[P,L]$. Hence the Lax equation is essentially a evolution equation for $q$
\be
 \partial_t q = p(q),
\ee
where $p(q)$ means a differential polynomial in $q$. We require the evolution equation to preserve 
gauge equivalence \footnote{This notion of preserving gauge equivalence has nothing to do with the 
gauge invariance of the flatness condition. The former is about the gauge transformation of the PDO or 
$q$ defined in the paper by Drinfeld and Sokolov, the latter is about the usual gauge transformation in 
field theory simultaneously acted on $a_x$ and $a_t$.}, that is, when starting with two initial conditions 
for $q$ which are connected by a gauge transformation, the two solutions should be also connected by 
a (time-dependent) gauge transformation at any time. The Lax equation preserving gauge equivalence is actually an evolution equation of gauge equivalent classes. Needless to say, we can choose 
representatives of some special form to specify the time evolution of the gauge equivalent classes. This motivates the definition of the canonical form of $L$, or $q$.We  denote the part of $q$ with weight $-i$ by 
$q_i$. In principle $q_i$ lies in the $N-|i|$ dimensional linear space spanned by $V^{|i|+1}_i,\ldots,V^N_i
$. By restricting $q_i$ to be in a one dimensional subspace, that is, a specific linear combination, we 
define a canonical form for $q$. For technical reasons, we also require the one dimensional subspace has a nonzero lowest weight projection. The name canonical form is justified by the following theorem, for 
any $q$ there is a unique gauge transformation to transform it into the canonical form, and the 
expression in the canonical form is unique. 
See Appendix \ref{app:B1} for a proof.  The  choice of the one dimensional subspaces that $q^\prime_i$ lie in defines the specific canonical form. Two choices 
are of particular importance in our discussion. The first one, we restrict $q^\prime_i$ to be lowest 
weight, if not an abuse of language, we call this the lowest weight canonical form. The second one, we restrict $q^\prime_i$ to be multiple of $e_{1,i+1}$, which we call the KdV canonical form. In the lowest weight canonical form,
\be
 q=\sum_{i=1}^N \alpha_i V^i_{-i+1},
\ee
the Lax equation $\frac{d}{dt}L=[P,L]$ gives us the flatness condition of Chern-Simons theory in the lowest weight gauge (by appropriately choosing $a_t$). In the KdV canonical form
\be
 q=-\sum_{i=1}^N u_i e_{1,i},
\ee
the Lax equation $\frac{d}{dt}L=[P,L]$ gives us KdV, as proved in the paper by Drinfeld and Sokolov. The evolution equation in the lowest weight canonical form and that in the KdV canonical form are just two special explicit forms of the same equation. There is a unique gauge transformation that transforms between these two canonical forms, which establish the one-to-one correspondence between Lifshitz Chern-Simons theory with $sl(N,\mathbb{R}),z$ and KdV with $n=N,m=z$, and explicitly the map from $\alpha_i$'s to $u_i$'s. From the relation
\be
 {\rm Tr} [P,L] = -{\rm Tr} [m_{-1},e] = 0,
\ee
it follows that the trace part of $L$ must be constant by the equation of motion. In the following  we set to be zero for simplicity. For example, we can set $\alpha_1=0$ for the $q$ in the lowest weight canonical form.

Now let's construct the conserved quantities from the Lax equation. In general, a matrix $A$ whose elements are power series in $\lambda$ (both positive and negative) can be uniquely expanded in the form
\be
 A=\sum_i a_i \Lambda^i,
\ee
where $a_i$'s are diagonal $\lambda$ independent matrices. Here $q$ is lower triangular, so it has the expansion $\sum_{i=0}^{N-1} d_i \Lambda^{-i}$, or
\be
 L=\partial_x + \Lambda + \sum_{i=0}^{N-1} d_i \Lambda^{-i}.
\ee
There is a similarity transformation to transform $L$ into a scalar coefficient form, that is, there is a formal series
\be
 T=E+\sum_{i=1}^\infty h_i \Lambda^{-i},
\ee
where $h_i$'s are diagonal matrices,
such that
\be
 L_0=T L T^{-1}=\partial_x + \Lambda + \sum_{i=0}^\infty f_i \Lambda^{-i},
\ee
where $f_i$'s are scalar functions, as opposed to matrices multiplied to the left. $T$ is determined up to multiplication by series of the form $E+\sum_{i=1}^\infty t_i \Lambda^i$ where $t_i$'s are scalar functions, and $f_i$'s are determined up to a total derivative. Most importantly 
\be
 q^i=\int f_i\ ,
\ee
are conserved by the Lax equation. See Appendix \ref{app:B2} for the proof.

The scalar coefficient form $L_0=\partial_x + \Lambda + \sum_{i=0}^\infty f_i \Lambda^{-i}$ not only gives us the conserved quantities, but also can help us to determine the form of the matrices that commute with $L$, and ultimately the form of $P$. Matrices that commute with $L_0$ must take the form $\sum_{i=-\infty}^n c_i \Lambda^i$ with $c_i$'s as constant coefficients, see Appendix \ref{app:B3} for a proof. Therefore matrices that commute with $L$ must have the form
\be
 M=T^{-1} \bigg(\sum_{i=-\infty}^n c_i \Lambda^i\bigg) T.
\ee
because $[M,L]=0$ is equivalent to $[TMT^{-1},L_0]=0$. Setting $P=M_+$, we get the consistent Lax equation $\frac{d}{dt}L=[P,L]$. Despite the simple appearance, several remarks about this equation are necessary. First, $T$ is the series that transforms $L$ into a form with scalar coefficients $L_0$ and it's in general a differential polynomial in $q$, hence $P$ is a differential polynomial in $q$ and so is the commutator $[P,L]$. Second, though $T$ has the indeterminacy of a multiplicative series $E+\sum_{i=1}^\infty t_i \Lambda^{-i}$ where $t_i$'s are scalar functions, $P$ is uniquely defined because $\sum_{i=-\infty}^n c_i \Lambda^i$ commute with this series. Last but the most important, this Lax equation preserves gauge equivalence,  a proof of this statement will be given in the Appendix \ref{app:B4}. 

As a evolution equation of gauge equivalent classes, the explicit form of the Lax equation $\frac{d}{dt}L=[P,L]$ is certainly not unique and different explicit forms correspond to choice of different representatives in gauge equivalent classes. We have the following theorem, if the difference between $P_1$ and $P_2$ is a negative weight matrix with no time or $\lambda$ dependence, then $\frac{d}{dt}L=[P_1,L]$ and $\frac{d}{dt}L=[P_2,L]$ give the same evolution equations of gauge equivalent classes. See Appendix \ref{app:B5} for a proof. Applying  this theorem, we can add a negative weight matrix both independent on time and $\lambda$ to $P$ without actually changing the evolution equation of gauge equivalent classes. We do need to do so when we want to obtain the Lax equation in certain canonical form, because the commutator $[P,L]$ is guaranteed to be negative weight, but not necessarily in the specific canonical form. The correction added to $P$ can be uniquely determined. The proof of this statement will be omitted because it's structurally the same as the proof of existence and uniqueness of the gauge transformation that transforms $L$ into a canonical form.

At last we have enough ingredients to explain how the integrable Lifshitz Chern-Simons theory for $sl(N,\mathbb{R})$ and $z$ emerges from the Drinfeld Sokolov formalism. First the Lax equation $\frac{d}{dt}L=[P,L]$ is equivalent to the flatness condition $\frac{d}{dt}D_x+[a_t,D_x]=0$ with the identification $a_x=V^2_1+q$ and $a_t=-\textit{Zero}(P)$. Second, the Lax equation, viewed as evolution equation of gauge equivalent classes, can be put in the lowest weight canonical form, which corresponds to lowest weight gauge choice in the Chern-Simons theory. Then, considering the Lifshitz exponent is $z$, we set $P= (T^{-1} \Lambda^z T)_+$ up to a multiplicative constant. At last we add a correction to $P$ to make $[P,L]$ lowest weight. From $P$ obtained in this way, $a_t=-\textit{Zero}(P)$ coincides with $a_t$ in "KdV  gauge" in our previous paper. If we choose the KdV canonical form for $L$, we get KdV hierarchy as proved in the paper by Drinfeld and Sokolov. The gauge transformation between the two canonical forms gives us the explicit map between the Lifshitz Chern-Simons theory and the KdV hierarchy. This map is $z$ independent simply because $z$ doesn't involve in the construction of gauge transformation between the two canonical forms.

\section{Discussion}
\setcounter{equation}{0}
\label{sec5}

In the present paper we showed that there is an explicit relation of the Chern-Simons Lifshitz theories and the integrable KdV hierarchy. This relation identifies the parameters $N$ and $z$ of the Chern Simons theory to the parameters  $n$ and $m$ of the KdV hierarchy. Consequently  the map  exists for all values of $N$. We discuss the status of the generalization to the infinite dimensional algebra $hs(\lambda)$.

  The fact that the equations of motion obey the scaling laws implies that the equation of motion, as well as the KdV map for a CS field $\alpha_i$, only contains finitely many terms since fields with a too large scaling dimension  cannot appear. Since the $hs(\lambda)$ truncates to $sl(N,\reals)$ and we adopted the normalization of our $sl(N,\reals)$ generators which is compatible with this truncation, for a finite number of fields the results for $sl(N,\reals)$ are mapped to the general $hs(\lambda)$ case by replacing $N\to \lambda$. It would nevertheless be interesting to see whether its possible to derive a closed form expression valid for all $a_i$. 
  
  The construction of the CS Lifshitz theory has a close relation to the construction of the asymptotically AdS theories which realize $W$ algebras, with many equations related by an exchange of lightcone coordinates $x^\pm $ with space and time $x,t$ (see \cite{Compere:2013gja} for the discussion of the  $SL(3,R)$  case). It would be interesting to see whether this relation can also be understood on the level of the conformal field theory, for some early discussion in this direction in the literature see   \cite{Mathieu:1991et,Kupershmidt:1989bf}.

In the present paper we have related the CS Lifshitz theory to the integrable KdV hierarchy. There exists a related and in some sense more universal integrable hierarchy the so called KP hierarchy \cite{kpeq}.
It would be interesting to investigate whether a relation of the CS Lifshitz theory for $hs(\lambda)$ to the KP hierarchy exists (see for possibly relevant work 
\cite{FigueroaO'Farrill:1992cv,FigueroaO'Farrill:1992uq,Khesin:1993ww,Khesin:1993ru,Yu:1991bk,Yu:1991ng}). We leave these interesting questions for future work.


\section*{Acknowledgements}

The work of M. Gutperle and Y. Li was supported in part by National Science Foundation grant PHY-13-13986. The work of M. Beccaria is supported 
 by  the  Russian Science Foundation grant 14-42-00047 and associated with Lebedev Institute.

\newpage

\appendix

\section{Conventions for gauge algebras}\label{appa}

In this appendix we collect our conventions for the $sl(N,\mathbb{R})$ and $hs(\lambda)$ algebras.
Recalling that for integer values of $\lambda$ the $hs(\lambda)$ algebra truncates to $sl(N,\mathbb{R})$ we use the same notation for the
generators of the two algebras.

\subsection{$sl(N,\mathbb{R})$ conventions}

In the fundamental representation the generators of the $sl(N,\mathbb{R})$ algebra are $N \times N$ matrices, labelled by two integers $s$, $m$ with $2\le s\le N$  and $|m|<s\le N$. 
All generators are built starting from 
the generators $\{V^{2}_{0}, V^{2}_{\pm 1}\}$ of the canonical $sl(2,\reals)$ subalgebra, whose non-zero 
matrix elements 
are given by (indices range from 1 to $N$)
\be
\left( V^2_0 \right)_{j,j}=\frac{N+1}{2}-j,\qquad
\left( V^2_1 \right)_{j+1,j} = - \sqrt{j(N-j)},\qquad
\left( V^2_{-1} \right)_{j,j+1} = \sqrt{j(N-j)}\,.
\ee
The other generators are obtained according to:
\be
V^s_m = (-1)^{s-m-1} \frac{(s + m -1)!}{(2 s - 2)!} \,
\underbrace{[V^2_{-1}, [V^2_{-1}, \dots, [V^2_{-1}, }_{s-m-1}\left( V^2_1 \right)^{s-1}]\dots].
\ee 
\subsection{$hs(\lambda)$ conventions}

The $hs(\lambda)$ algebra is spanned by the infinite set of  generators  $V^s_m$, $s=1,2,3,\ldots$ and $m=-s+1,-s+2,\ldots,s-1$. 
The associative lone star product is defined as
\be
 V^s_m * V^t_n = \frac{1}{2} \sum_{u=1}^{s+t-|s-t|-1} g^{st}_u(m,n,\lambda) V^{s+t-u}_{m+n}.
\ee
The structure constants of the $hs(\lambda)$ algebra were defined in
\cite{Pope:1989sr} and can be represented as follows
\be\label{gdefa}
g_u^{st}(m,n;\lambda) = {q^{u-2}\over2(u-1)!}\phi_u^{st}(\lambda)N_u^{st}(m,n),
\ee
where $q$ is a normalization constant which can be eliminated by a rescaling of the generators; we choose $q=1/4$ to agree with the literature. The other terms in (\ref{gdefa}) are given by 
\bea
 N_u^{st}(m,n) &=& \sum_{k=0}^{u-1}(-1)^k
\left( 
\begin{array}{c}
     u-1  \\
k
\end{array}
\right)
[s-1+m]_{u-1-k}[s-1-m]_k[t-1+n]_k[t-1-n]_{u-1-k},\nonumber\\
\phi_u^{st}(\lambda) &= &\ _4F_3\left[
\begin{array}{cccc}
 \half + \lambda  &   \half - \lambda &{2-u\over 2}   &{1-u\over 2}\\
 {3\over 2}-s  &  {3\over 2} -t &  \half + s+t-u &\\  
\end{array} \Bigg| 1
\right].
\eea
The descending Pochhammer symbol $[a]_n$ is defined as
\be
[a]_n  = a(a-1)...(a-n+1)~,
\ee
and the commutator is defined as
\be
 [V^s_m, V^t_n]=V^s_m * V^t_n - V^t_n * V^s_m.
\ee
$V^1_0$ is the unit element. The trace of a $hs(\lambda)$ element is defined as the coefficient of $V^1_0$ up to a multiplicative constant ${\rm tr}(V^1_0)$. When $\lambda=N$ 
integer, $hs(\lambda)$ is truncated to $sl(N,\reals)$. That means, we can consistently set $V^s_m$ to be zero if $s>N$, and the remaining elements can be 
identified with the $sl(N,\reals)$ generators defined above; the star product becomes the usual matrix multiplication and the trace the usual matrix trace.


\section{Proofs of statements used  in the Drinfeld Sokolov formalism}\label{appb}

In this part of appendix we give the proofs to the theorems used in Drinfeld Sokolov formalism. Most of them are essentially contained in the original paper by Drinfeld and Sokolov. However, the original paper is a little bit condensed, so we add details to the proofs to make them easier to follow.

\subsection{Gauge transformation of PDOs}
\label{app:B1}

Here we give the proof of the following statement:  For any $q$ and any canonical form, there exist a unique gauge transformation $S$ to transform $q$ into $q^\prime=S^{-1} V^2_1 S - V^2_1 + S^{-1} \partial_x S$ in the canonical form chosen. 

The proof proceeds as follows: We rewrite the gauge transformation as
\be
 S q^\prime = q S + [V^2_1,S] + \partial_x S
\ee
and then by comparing the weight $-i$ part we get
\be
 \sum_{j=0}^i S_{i-j}q^\prime_j=\sum_{j=0}^i q_j S_{i-j} + [V^2_1,S_{i+1}] + \partial_x S_i
\ee
which holds for all $i$'s. Using the fact $S_0$ is the identity matrix $E$, we put it in a recursive form
\be
 q^\prime_i-[V^2_1,S_{i+1}]= q_i + \partial_x S_i-\sum_{j=0}^{i-1}S_{i-j}q^\prime_j+\sum_{j=0}^{i-1} q_j S_{i-j}.
\ee
Given $q$, and suppose $q^\prime_j$ and $S_{j+1}$ are known for all $j<i$, from the lowest weight projection of the right hand side we can find $q^\prime_i$ if we restrict it to be in a one dimensional subspace of weight $-i$ elements which has nonzero lowest weight projection. Then $S_{i+1}$ is also determined by equating non lowest weight terms on both sides. The initial conditions, needless to say, are $q^\prime_0=q_0$ and $S_0=E$.

\subsection{ Scalar coefficient form and conserved quantities}
\label{app:B2}

Here we proof the following statement: For generic $L$, there is a formal series
\be
 T=E+\sum_{i=1}^\infty h_i \Lambda^{-i},
\ee
where $h_i$'s are diagonal matrices,
such that
\be
 L_0=T L T^{-1}=\partial_x + \Lambda + \sum_{i=0}^\infty f_i \Lambda^{-i},
\ee
where $f_i$'s are scalar functions. $T$ is determined up to multiplication by series of the form $E+\sum_{i=1}^\infty t_i \Lambda^i$ where $t_i$'s are scalar functions, and $f_i$'s are determined up to a total derivative. Furthermore $q^i=\int f_i$  are conserved by the Lax equation.

The proof proceeds as follows:  By equating the coefficients of the same powers of $\Lambda$ in the equality $T L=L_0 T$ we get
\be
 d_i+h_{i+1}+\sum_{j=0}^{i-1}h_{i-j}d_j^{\sigma^{-(i-j)}}=f_i E+\partial_x h_i+h_{i+1}^\sigma+\sum_{j=1}^i f_{i-j}h_j^{\sigma^{-(i-j)}}.
\ee
Here the notation $A^{\sigma^i}$ means $\Lambda^i A \Lambda^{-i}$, which is $i$ times cyclic permutation of the diagonal elements for a diagonal matrix $A$. For example if $A=Diag\{a_1,a_2,a_3,a_4\}$ then $A^\sigma=Diag\{a_2,a_3,a_4,a_1\}$.
We rewrite the equation above as
\be
 h_{i+1}-h_{i+1}^\sigma-f_i E=-d_i+\partial_x h_i-\sum_{j=0}^{i-1}h_{i-j}d_j^{\sigma^{-(i-j)}}+\sum_{j=1}^i f_{i-j}h_j^{\sigma^{-(i-j)}}.
\ee
$f_i$ is obtained by taking the trace on both sides, then $h_{i+1}$ is determined up to an additive multiple of identity. Now suppose $T^\prime$ transforms $L$ to
\be
 L_0^\prime=T^\prime L T^{\prime -1}=\partial_x + \Lambda + \sum_{i=0}^\infty f^\prime_i \Lambda^{-i}.
\ee
Define $T T^{\prime -1}=A=E+\sum_{i=1}^\infty a_i \Lambda^i$ where $a_i$'s are diagonal matrices. We have $A^{-1} L_0 A=L_0^\prime$ or $L_0 A = A L_0^\prime$. By equating the coefficients of the same power in $\Lambda$ we get
\be
 a_{i+1}-a_{i+1}^\sigma + f_i^\prime E - f_i E = \partial_x a_i + \sum_{j=0}^{i-1} f_i a^{\sigma^{-i}}_{i-j} - \sum_{j=0}^{i-1} f_i^\prime a_{i-j}
\ee
with the initial conditions
\bea
 a_1-a_1^\sigma + f_0^\prime E - f_0 E&=&0, \nonumber \\
 a_2-a_2^\sigma + f_1^\prime E - f_1 E&=&\partial_x a_1.
\eea
From this recursive formula it's easy to see $a_i-a_i^\sigma=0$ for all $i$, that is $a_i$'s are all multiples of identity, say, $a_i=t_i E$. Plug this back into the recursive formula we have
\be
 f_i^\prime-f_i = \partial_x t_i - \sum_{j=0}^{i-1} t_{i-j} (f_j^\prime-f_j)
\ee
with the initial condition
\bea
 f_0^\prime-f_0 &=& 0, \nonumber \\
 f_1^\prime-f_1 &=& \partial_x t_1.
\eea
One can prove by induction that $f_i^\prime-f_i$ is a total derivative.

The evolution equation of $L_0$ is
\be
 \frac{d}{dt}L_0=[P_0,L_0],
\ee
where $P_0=\frac{dT}{dt}T^{-1}+TPT^{-1}$. Expand $P_0$ as $\sum_{i=-\infty}^n p_i \Lambda^i$, then the Lax equation above gives us
\bea
 0&=&p_n-p_n^\sigma, \nonumber \\
 0&=&-\partial_x p_i+p_{i-1}-p_{i-1}^\sigma+\sum_{j=i}^n f_{j-i} (p_j-p_j^{\sigma^{j-i}}),\quad  0<i \leq n, \nonumber \\
 \dot{f}_{-i}&=&-\partial_x p_i+p_{i-1}-p_{i-1}^\sigma+\sum_{j=i}^n f_{j-i} (p_j-p_j^{\sigma^{j-i}}), \quad i \leq 0.
\eea
This recursive formula demands all $p_i$'s to be multiples of identity. From this, in turn, the commutator simplifies to $-\partial_x P_0$, hence $\dot{f_i}$'s are equal to total derivatives and $\int f_i$'s are conserved.

\subsection{ Matrices that commute with $L_0$}
\label{app:B3}

Here we would like to show that  All matrices that commute with $L_0=\partial_x + \Lambda + \sum_{i=0}^\infty f_i \Lambda^{-i}$ have the form $\sum_{i=-\infty}^n c_i \Lambda^i$ with $c_i$'s as constant coefficients.

This follows from letting $M=\sum_{i=-\infty}^n m_i \Lambda^i$ be a matrix commuting with $L_0$. By equating coefficients of the same power in $\Lambda$ in the equation $M L_0=L_0 M$ we get
\bea
 m_n-m_n^\sigma &=& 0, \nonumber \\
 -\partial_x m_i+m_{i-1}-m_{i-1}^\sigma+\sum_{j=i}^n f_{j-i} (m_j-m_j^{\sigma^{j-i}})&=&0, \quad i \leq n.
\eea
Therefore all $m_i$'s are constants times identity matrix.

\subsection{The Lax equation preserves gauge equivalence}
\label{app:B4}

In this subsection we prove the statement that by choosing $P=(T^{-1} (\sum_{i=-\infty}^n c_i \Lambda^i) T)_+$ the Lax equation preserves gauge equivalence. 

This can be shown as follows:  It suffices to prove if $L$ satisfies the Lax equation, then so does $L^\prime=S^{-1} L S$ where $S$ is a gauge transformation matrix that only depends on $x$. In other words $\partial_t q = p(q)$ implies $\partial_t q^\prime = p(q^\prime)$. Using the original Lax equation, it's straightforward to get
\be
 \frac{d}{dt} L^\prime = [S^{-1} P S,L^\prime].
\ee
So we want $S^{-1} P S = P^\prime$, which means, $S^{-1} P S$ is the same differential polynomial in $q^\prime$ as $P$ in $q$. Explicitly we have
\be
 S^{-1} P S = S^{-1} (T^{-1} (\sum_{i=-\infty}^n c_i \Lambda^i) T)_+ S = ( (TS)^{-1} (\sum_{i=-\infty}^n c_i \Lambda^i) (TS))_+.
\ee
Suppose $T^\prime$ transforms $L^\prime$ into the form of scalar coefficients, that is $T^\prime L^\prime T^{\prime -1}= L_0^\prime$, so $T^\prime$ is the same differential polynomial in $q^\prime$ as $T$ in $q$. Plug in $L^\prime=S^{-1} L S$ we get $(T^\prime S^{-1}) L (T^\prime S^{-1})^{-1} = L_0^\prime = L_0 = TLT^{-1}$. Hence $T^\prime S^{-1}=T$ or $TS=T^\prime$, and at last we get 
\be
 S^{-1} P S = ( T^{\prime -1} (\sum_{i=-\infty}^n c_i \Lambda^i) T^\prime)_+=P^\prime.
\ee

\subsection{Equivalent evolution equations of gauge equivalent classes}
\label{app:B5}

We want to prove the following statement: Given that the difference between $P_1$ and $P_2$ is a negative weight matrix with no time or $\lambda$ dependence, then $\frac{d}{dt}L=[P_1,L]$ and $\frac{d}{dt}L=[P_2,L]$ give the same evolution equations of gauge equivalent classes. 

The proof proceeds as follows: Let's $R$ denote the ring of scalar differential polynomials in $q$ which are invariant under gauge transformation. For any $f \in R$ the time derivative of $f$ by the Lax equation also belongs to $R$, and the form of time derivatives of all $f \in R$ uniquely specify the evolution equation of gauge equivalent classes. Now for any  $f \in R$, let $g$ be the difference of the time derivative of $f$ by the above two Lax equations, then $g$ is actually the time derivative of $f$ by the Lax equation $\frac{d}{dt}L=[P_1-P_2,L]$. Formally
\be
 g(L)=\frac{d}{dt} f({\cal L}(t)) |_{t=0},
\ee
where ${\cal L}(t)$ satisfies
\bea
 {\cal L}(0)&=&L,\nonumber \\
 \frac{d}{dt} {\cal L}(t) |_{t=0} &=& [P_1-P_2,L].
\eea
Apparently ${\cal L}(t)= S L S^{-1}$ where $S=E+t(P_1-P_2)$ satisfies these conditions, and its time evolution is just a gauge transformation. Therefore we have $g=0$ because $g \in R$.

\section{Explicit results for various $N$ and $z$}
\label{appkdv}

In this appendix we collect  explicit results for several pairs $(N, z)$. For each $N$, we list the 
$z$-independent CS-KdV map and, for various $z$, the explicit KdV and CS equations of motion. 
Due to the length of the equations we don't write all the cases for $N=6$ and $N=7$, 
limiting the presentation to the first values of $z$ (up to $z=4, 3$ respectively). \footnote{Additional data are available 
from the authors upon request.}

\bigskip
\noindent
\underline{\boldmath$N=3$}

\medskip
\noindent
CS-KdV map:
\bea
 u_2 &=& \,\, 4  \,\alpha_2  ,  \notag \\ 
 u_3 &= & \,\, 2  \,\alpha_2' -4  \,\alpha_3  .
 \eea
KdV equations of motion at $z=2$:
\bea
  \dot u_2 &=&  2 \,u_3' -\,u_2'',   \notag \\ 
  \dot u_3 &=&  -\frac{2}{3} \,u_2  \,u_2' +\,u_3'' -\frac{2}{3} \,u_2'''.
 \eea
CS equations of motion at $z=2$:
 \bea
 \dot \alpha_2 &=& -2  \,\alpha_3'  , \notag \\ 
 \dot \alpha_3 &=& \frac{8}{3}  \,\alpha_2   \,\alpha_2' +\frac{1}{6}  \,\alpha_2'''.
 \eea


\bigskip
\noindent
\underline{\boldmath$N=4$}

\medskip
\noindent
CS-KdV map:
\bea
 u_2 &=& \,\, 10  \,\alpha_2  ,  \notag \\ 
 u_3 &=& \,\, 10  \,\alpha_2' -24  \,\alpha_3  ,   \\ 
 u_4 &=& \,\, -12  \,\alpha_3' +3  \,\alpha_2'' +9  \,\alpha_2 {}^2+36  \,\alpha_4 \notag .
 \eea
KdV equations of motion at $z=2$:
\bea
  \dot u_2 & =&  2 \,u_3' -2 \,u_2'',   \notag \\ 
  \dot u_3 & =& -\,u_2  \,u_2' +2 \,u_4' +\,u_3'' -2 \,u_2''',    \\ 
  \dot u_4 & =& -\frac{1}{2} \,u_3  \,u_2' -\frac{1}{2} \,u_2  \,u_2'' +\,u_4'' -\frac{1}{2} \,u_2^{\text{(4)}}\notag.  
 \eea
CS equations of motion at $z=2$:
 \bea
 \dot \alpha_2 &= & -\frac{24}{5}  \,\alpha_3'  , \notag \\ 
 \dot \alpha_3 &= & \frac{8}{3}  \,\alpha_2   \,\alpha_2' -3  \,\alpha_4' +\frac{1}{6}  \,\alpha_2'''  ,  \\ 
 \dot \alpha_4 &= & \frac{10}{3}  \,\alpha_3   \,\alpha_2' +\frac{12}{5}  \,\alpha_2   \,\alpha_3' +\frac{1}{15}  \,\alpha_3'''\notag.
  \eea
KdV equations of motion at $z=3$:
\bea
  \dot u_2 & =&  -\frac{3}{4} \,u_2  \,u_2' +3 \,u_4' -\frac{3}{2} \,u_3'' +\frac{1}{4} \,u_2''',  \notag \\ 
  \dot u_3 & =&  -\frac{3}{4} \,u_3  \,u_2' -\frac{3}{4} \,u_2  \,u_3' +3 \,u_4'' -2 \,u_3''' +\frac{3}{4} \,u_2^{\text{(4)}},    \\ 
  \dot u_4 & =&  -\frac{3}{4} \,u_3  \,u_3' +\frac{3}{4} \,u_2  \,u_4' +\frac{3}{8} \,u_3  \,u_2'' -\frac{3}{4} \,u_2  \,u_3'' +\frac{3}{8} \,u_2 
   \,u_2''' +\,u_4''' -\frac{3}{4} \,u_3^{\text{(4)}} +\frac{3}{8} \,u_2^{\text{(5)}}\notag.
 \eea
CS equations of motion at  $z=3$ :
 \bea
 \dot \alpha_2 &= & -\frac{21}{10}  \,\alpha_2   \,\alpha_2' +\frac{54}{5}  \,\alpha_4' -\frac{7}{20}  \,\alpha_2'''  , \notag \\ 
 \dot \alpha_3 &= & -\frac{15}{2}  \,\alpha_3   \,\alpha_2' -\frac{15}{2}  \,\alpha_2   \,\alpha_3' -\frac{1}{2}  \,\alpha_3'''  ,  \\ 
 \dot \alpha_4 &= & \frac{59}{60}  \,\alpha_2'   \,\alpha_2'' +\frac{24}{5}  \,\alpha_2 {}^2  \,\alpha_2' +\frac{21}{10}  \,\alpha_2   \,\alpha_4' -12  \,\alpha_3  \,\alpha
   _3' +\frac{13}{30}  \,\alpha_2   \,\alpha_2''' +\frac{1}{10}  \,\alpha_4''' +\frac{1}{120}  \,\alpha_2^{\text{(5)}} \notag.
 \eea

\bigskip
\noindent
\underline{\boldmath$N=5$}

\medskip
\noindent
CS-KdV map:
\bea
 u_2  &=  & \,\, 20  \,\alpha_2  ,  \notag \\ 
 u_3  &=  & \,\, 30  \,\alpha_2' -84  \,\alpha_3  ,  \notag \\ 
 u_4  &=  & \,\, -84  \,\alpha_3' +18  \,\alpha_2'' +64  \,\alpha_2 {}^2+288  \,\alpha_4  ,   \\ 
 u_5 &=  & \,\, 64  \,\alpha_2   \,\alpha_2' +144  \,\alpha_4' -24  \,\alpha_3'' +4  \,\alpha_2''' -192  \,\alpha_2   \,\alpha_3 -576  \,\alpha_5 \notag .
 \eea
KdV equations of motion at $z=2$:
\bea
  \dot u_2 & =&  2 \,u_3' -3 \,u_2'',   \notag \\ 
  \dot u_3 & =&  -\frac{6}{5} \,u_2  \,u_2' +2 \,u_4' +\,u_3'' -4 \,u_2''' ,  \notag \\ 
  \dot u_4 & =&  -\frac{4}{5} \,u_3  \,u_2' +2 \,u_5' -\frac{6}{5} \,u_2  \,u_2'' +\,u_4'' -2 \,u_2^{\text{(4)}},    \\ 
  \dot u_5 & =&  -\frac{2}{5} \,u_4  \,u_2' -\frac{2}{5} \,u_3  \,u_2'' +\,u_5'' -\frac{2}{5} \,u_2  \,u_2''' -\frac{2}{5} \,u_2^{\text{(5)}}\notag.
   \eea
CS equations of motion at $z=2$:   
\bea
 \dot \alpha_2 &= & -\frac{42}{5}  \,\alpha_3'  , \notag \\ 
 \dot \alpha_3 &= & \frac{8}{3}  \,\alpha_2   \,\alpha_2' -\frac{48}{7}  \,\alpha_4' +\frac{1}{6}  \,\alpha_2'''  , \notag \\ 
 \dot \alpha_4 &= & \frac{10}{3}  \,\alpha_3   \,\alpha_2' +\frac{12}{5}  \,\alpha_2   \,\alpha_3' -4  \,\alpha_5' +\frac{1}{15}  \,\alpha_3'''  ,  \\ 
 \dot \alpha_5 &= & \frac{14}{5}  \,\alpha_3   \,\alpha_3' +4  \,\alpha_4   \,\alpha_2' +\frac{16}{7}  \,\alpha_2   \,\alpha_4' +\frac{1}{28}  \,\alpha_4'''\notag.
  \eea
KdV equations of motion at $z=3$:
\bea
  \dot u_2 & =&  -\frac{6}{5} \,u_2  \,u_2' +3 \,u_4' -3 \,u_3'' +\,u_2''',   \notag \\ 
  \dot u_3 & =&  -\frac{6}{5} \,u_3  \,u_2' -\frac{6}{5} \,u_2  \,u_3' +3 \,u_5' +3 \,u_4'' -5 \,u_3''' +3 \,u_2^{\text{(4)}},   \notag \\ 
  \dot u_4 & =&  -\frac{6}{5} \,u_3  \,u_3' -\frac{3}{5} \,u_4  \,u_2' +\frac{3}{5} \,u_2  \,u_4' +\frac{3}{5} \,u_3  \,u_2'' -\frac{9}{5} \,u_2  \,u_3'' +3
   \,u_5'' +\frac{6}{5} \,u_2  \,u_2''' +  \notag \\  && 
   \,u_4''' -3 \,u_3^{\text{(4)}} +\frac{12}{5} \,u_2^{\text{(5)}},    \\ 
  \dot u_5 & =&  -\frac{3}{5} \,u_4  \,u_3' +\frac{3}{5} \,u_2  \,u_5' -\frac{3}{5} \,u_3  \,u_3'' +\frac{3}{5} \,u_4  \,u_2'' +\frac{3}{5} \,u_3  \,u_2''' -\frac{3}{5} \,u_2 
   \,u_3''' +\,u_5''' + \notag \\ && 
   \frac{3}{5} \,u_2  \,u_2^{\text{(4)}} -\frac{3}{5} \,u_3^{\text{(5)}} +\frac{3}{5} \,u_2^{\text{(6)}}\notag.
\eea
CS equations of motion at $z=3$:
\bea
 \dot \alpha_2 &= & -\frac{24}{5}  \,\alpha_2   \,\alpha_2' +\frac{216}{5}  \,\alpha_4' -\frac{4}{5}  \,\alpha_2'''  , \notag \\ 
 \dot \alpha_3 &= & -\frac{120}{7}  \,\alpha_3   \,\alpha_2' -\frac{120}{7}  \,\alpha_2   \,\alpha_3' +\frac{144}{7}  \,\alpha_5' -\frac{8}{7}  \,\alpha_3'''  , \notag \\ 
 \dot \alpha_4 &= & \frac{59}{60}  \,\alpha_2'   \,\alpha_2'' +\frac{24}{5}  \,\alpha_2 {}^2  \,\alpha_2' -\frac{36}{5}  \,\alpha_2   \,\alpha_4' -\frac{147}{5}  \,\alpha_3  \,\alpha
   _3' -12  \,\alpha_4   \,\alpha_2' +\frac{13}{30}  \,\alpha_2   \,\alpha_2''' -
     \notag \\  &&  
    \frac{1}{5}  \,\alpha_4''' +\frac{1}{120}  \,\alpha_2^{\text{(5)}}  ,  \\ 
 \dot \alpha_5 &= & \frac{97}{140}  \,\alpha_3'   \,\alpha_2'' +\frac{29}{56}  \,\alpha_2'   \,\alpha_3'' +\frac{144}{35}  \,\alpha_2 {}^2  \,\alpha_3' +\frac{396}{35}  \,\alpha_3 
    \,\alpha_2   \,\alpha_2' +\frac{36}{7}  \,\alpha_2   \,\alpha_5' -\frac{126}{5}  \,\alpha_4   \,\alpha_3' -\frac{72}{5}  \,\alpha_3   \,\alpha_4'  +
 \notag \\  &&  
    \frac{5}{28}  \,\alpha_2  \,\alpha
   _3''' +\frac{123}{280}  \,\alpha_3   \,\alpha_2''' +\frac{1}{7}  \,\alpha_5''' +\frac{1}{560}  \,\alpha_3^{\text{(5)}}\notag .
 \eea
KdV equations of motion at $z=4$:
\bea
  \dot u_2 & =&  \frac{6}{5} \,u_2' {}^2-\frac{4}{5} \,u_3  \,u_2' -\frac{4}{5} \,u_2  \,u_3' +4 \,u_5' +\frac{6}{5} \,u_2  \,u_2'' -2 \,u_4'' +\,u_2^{\text{(4)}},
    \notag \\ 
  \dot u_3 & =&  \frac{24}{5} \,u_2'  \,u_2'' +\frac{12}{25} \,u_2 {}^2 \,u_2' -\frac{4}{5} \,u_2  \,u_4' -\frac{2}{5} \,u_2'  \,u_3' -\frac{4}{5} \,u_3 
   \,u_3' -\frac{4}{5} \,u_4  \,u_2' -\frac{2}{5} \,u_2  \,u_3'' + 
    \notag \\ &&  6 \,u_5'' +2 \,u_2  \,u_2''' -4 \,u_4''' +\,u_3^{\text{(4)}} +\frac{6}{5} \,u_2^{\text{(5)}},   \notag \\ 
  \dot u_4 & =&  \frac{16}{5} \,u_2'  \,u_2''' +\frac{12}{25} \,u_2  \,u_2' {}^2+\frac{8}{25} \,u_3  \,u_2  \,u_2' +\frac{8}{5} \,u_2  \,u_5' -\frac{4}{5} \,u_4 
   \,u_3' -\frac{2}{5} \,u_2'  \,u_4' - \notag \\  && \frac{4}{5} \,u_3  \,u_4' +
   \frac{12}{25} \,u_2 {}^2 \,u_2'' -\frac{8}{5} \,u_2  \,u_4'' +\frac{12}{5}
   \,u_2'' {}^2+\frac{2}{5} \,u_4  \,u_2'' +\frac{2}{5} \,u_2  \,u_3''' +\frac{2}{5} \,u_3  \,u_2''' +
     \notag \\  && 
   4 \,u_5''' +\frac{6}{5} \,u_2  \,u_2^{\text{(4)}} -3 \,u_4^{\text{(4)}} + 
   \frac{6}{5} \,u_3^{\text{(5)}} +\frac{2}{5} \,u_2^{\text{(6)}},    \\ 
  \dot u_5 & =&  \frac{12}{25} \,u_2  \,u_2'  \,u_2'' +\frac{4}{5} \,u_2'  \,u_2^{\text{(4)}} +\frac{4}{25} \,u_4  \,u_2  \,u_2' +\frac{4}{25} \,u_3  \,u_2' {}^2-\frac{4}{5}
   \,u_4  \,u_4' -\frac{2}{5} \,u_2'  \,u_5' +\frac{4}{5} \,u_3  \,u_5' +
     \notag \\  && 
   \frac{8}{5} \,u_2''  \,u_2''' +\frac{4}{25} \,u_3  \,u_2 
   \,u_2'' +\frac{4}{5} \,u_2  \,u_5'' +\frac{2}{5} \,u_4  \,u_3'' -\frac{4}{5} \,u_3  \,u_4'' +\frac{4}{25} \,u_2 {}^2 \,u_2''' -\frac{4}{5} \,u_2 
   \,u_4''' +
     \notag \\  && 
   \frac{2}{5} \,u_3  \,u_3''' +\frac{2}{5} \,u_2  \,u_3^{\text{(4)}} +\,u_5^{\text{(4)}} +\frac{4}{25} \,u_2  \,u_2^{\text{(5)}} -\frac{4}{5} \,u_4^{\text{(5)}} +\frac{2}{5} \,u_3^{\text{(6)}} \notag.
 \eea
CS equations of motion at $z=4$
\bea
 \dot \alpha_2 &= & \frac{144}{5}  \,\alpha_3   \,\alpha_2' +\frac{144}{5}  \,\alpha_2   \,\alpha_3' -\frac{576}{5}  \,\alpha_5' +\frac{18}{5}  \,\alpha_3'''  , \notag \\ 
 \dot \alpha_3 &= & -\frac{24}{7}  \,\alpha_2'   \,\alpha_2'' -\frac{64}{7}  \,\alpha_2 {}^2  \,\alpha_2' +\frac{384}{7}  \,\alpha_2   \,\alpha_4' +\frac{336}{5}  \,\alpha_3  \,\alpha
   _3' +\frac{384}{7}  \,\alpha_4   \,\alpha_2' -
     \notag \\  &&  
   \frac{12}{7}  \,\alpha_2   \,\alpha_2''' +\frac{24}{7}  \,\alpha_4''' -\frac{1}{14}  \,\alpha_2^{\text{(5)}}  , \notag \\ 
 \dot \alpha_4 &= & -\frac{68}{15}  \,\alpha_3'   \,\alpha_2'' -\frac{61}{15}  \,\alpha_2'   \,\alpha_3'' -\frac{96}{5}  \,\alpha_2 {}^2  \,\alpha_3' -\frac{208}{5}  \,\alpha_3  \,\alpha
   _2   \,\alpha_2' -\frac{64}{5}  \,\alpha_2   \,\alpha_5' +\frac{336}{5}  \,\alpha_4   \,\alpha_3' +
      \notag \\  &&  
     \frac{336}{5}  \,\alpha_3   \,\alpha_4' 
   -\frac{26}{15}  \,\alpha_2  \,\alpha
   _3''' -\frac{13}{5}  \,\alpha_3   \,\alpha_2''' -\frac{4}{5}  \,\alpha_5''' -\frac{1}{30}  \,\alpha_3^{\text{(5)}}  ,  \\ 
 \dot \alpha_5 &= & \frac{1108}{315}  \,\alpha_2   \,\alpha_2'   \,\alpha_2'' -\frac{7}{2}  \,\alpha_3'   \,\alpha_3'' +\frac{12}{35}  \,\alpha_4'   \,\alpha_2'' +\frac{8}{7}
    \,\alpha_2'   \,\alpha_4'' +\frac{13}{168}  \,\alpha_2'   \,\alpha_2^{\text{(4)}} +\frac{256}{35}  \,\alpha_2 {}^3  \,\alpha_2' +
      \notag \\  &&  
    \frac{256}{35}  \,\alpha_2 {}^2 \,\alpha
   _4' -\frac{272}{5}  \,\alpha_3   \,\alpha_2   \,\alpha_3' 
   +\frac{32}{35}  \,\alpha_4   \,\alpha_2   \,\alpha_2' +\frac{62}{63}  \,\alpha_2' {}^3-32  \,\alpha_3 {}^2 \,\alpha
   _2' +\frac{576}{5}  \,\alpha_4   \,\alpha_4' -
     \notag \\  &&  
   \frac{144}{5}  \,\alpha_3   \,\alpha_5' +\frac{47}{360}  \,\alpha_2''   \,\alpha_2''' +\frac{244}{315}  \,\alpha_2 {}^2 \,\alpha
   _2''' 
   +\frac{4}{7}  \,\alpha_2   \,\alpha_4''' -\frac{19}{10}  \,\alpha_3   \,\alpha_3''' -\frac{38}{35}  \,\alpha_4   \,\alpha_2''' +
     \notag \\  &&  
   \frac{29  \,\alpha_2   \,\alpha_2^{\text{(5)}} }{1260}+\frac{1}{140}
    \,\alpha_4^{\text{(5)}} +\frac{ \,\alpha_2^{\text{(7)}} }{5040} \notag.
 \eea


\bigskip
\noindent
\underline{\boldmath$N=6$}

\medskip
\noindent
CS-KdV map:
\bea
 u_2 &=  & \,\, 35  \,\alpha_2  ,  \notag \\ 
 u_3 &=  & \,\, 70  \,\alpha_2' -224  \,\alpha_3  ,  \notag \\ 
 u_4 &=  & \,\, -336  \,\alpha_3' +63  \,\alpha_2'' +259  \,\alpha_2 {}^2+1296  \,\alpha_4  ,   \\ 
 u_5 &=  & \,\, 518  \,\alpha_2   \,\alpha_2' +1296  \,\alpha_4' -192  \,\alpha_3'' +28  \,\alpha_2''' -1760  \,\alpha_2   \,\alpha_3 -5760  \,\alpha_5  ,  \notag \\ 
 u_6 &=  & \,\, -880  \,\alpha_2   \,\alpha_3' +130  \,\alpha_2' {}^2-880  \,\alpha_3   \,\alpha_2' -2880  \,\alpha_5' +155  \,\alpha_2   \,\alpha_2'' +360  \,\alpha_4'' -40 \,\alpha
   _3''' +5  \,\alpha_2^{\text{(4)}} 
       \notag \\  &&  
   +225  \,\alpha_2 {}^3+3600  \,\alpha_4   \,\alpha_2 +1600  \,\alpha_3 {}^2+14400  \,\alpha_6 \notag  .
 \eea
KdV equations of motion at $z=2$:
\bea
  \dot u_2 & =&  2 \,u_3' -4 \,u_2''   \notag \\ 
  \dot u_3 & =&  -\frac{4}{3} \,u_2  \,u_2' +2 \,u_4' +\,u_3'' -\frac{20}{3} \,u_2''',   \notag \\ 
  \dot u_4 & =&  -\,u_3  \,u_2' +2 \,u_5' -2 \,u_2  \,u_2'' +\,u_4'' -5 \,u_2^{\text{(4)}},    \\ 
  \dot u_5 & =&  -\frac{2}{3} \,u_4  \,u_2' +2 \text{u6}' -\,u_3  \,u_2'' +\,u_5'' -\frac{4}{3} \,u_2  \,u_2''' -2 \,u_2^{\text{(5)}},  \notag \\ 
  \dot u_6 & =&  -\frac{1}{3} \,u_5  \,u_2' -\frac{1}{3} \,u_4  \,u_2'' +\text{u6}'' -\frac{1}{3} \,u_3  \,u_2''' -\frac{1}{3} \,u_2  \,u_2^{\text{(4)}} -\frac{1}{3} \,u_2^{\text{(6)}}\notag.
 \eea
CS equations of motion at $z=2$:
 \bea
 \dot \alpha_2 &= & -\frac{64}{5}  \,\alpha_3'  , \notag \\ 
 \dot \alpha_3 &= & \frac{8}{3}  \,\alpha_2   \,\alpha_2' -\frac{81}{7}  \,\alpha_4' +\frac{1}{6}  \,\alpha_2'''  , \notag \\ 
 \dot \alpha_4 &= & \frac{10}{3}  \,\alpha_3   \,\alpha_2' +\frac{12}{5}  \,\alpha_2   \,\alpha_3' -\frac{80}{9}  \,\alpha_5' +\frac{1}{15}  \,\alpha_3'''  ,  \\ 
 \dot \alpha_5 &= & \frac{14}{5}  \,\alpha_3   \,\alpha_3' +4  \,\alpha_4   \,\alpha_2' +\frac{16}{7}  \,\alpha_2   \,\alpha_4' -5  \,\alpha_6' +\frac{1}{28}  \,\alpha_4'''  , \notag \\ 
 \dot \alpha_6 &= & \frac{16}{5}  \,\alpha_4   \,\alpha_3' +\frac{18}{7}  \,\alpha_3   \,\alpha_4' +\frac{14}{3}  \,\alpha_5   \,\alpha_2' +\frac{20}{9}  \,\alpha_2   \,\alpha_5' +\frac{1}{45}
    \,\alpha_5''' \notag .
 \eea
KdV equations of motion at $z=3$:
\bea
  \dot u_2 & =&  -\frac{3}{2} \,u_2  \,u_2' +3 \,u_4' -\frac{9}{2} \,u_3'' +\frac{9}{4} \,u_2''',   \notag \\ 
  \dot u_3 & =&  -\frac{3}{2} \,u_3  \,u_2' -\frac{3}{2} \,u_2  \,u_3' +3 \,u_5' +3 \,u_4'' -9 \,u_3''' +\frac{15}{2} \,u_2^{\text{(4)}},   \notag \\ 
  \dot u_4 & =&  -\frac{3}{2} \,u_3  \,u_3' -\,u_4  \,u_2' +\frac{1}{2} \,u_2  \,u_4' +3 \text{u6}' +\frac{3}{4} \,u_3  \,u_2'' -3 \,u_2  \,u_3'' +3
   \,u_5'' +\frac{5}{2} \,u_2  \,u_2''' +\,u_4'''- 
       \notag \\  && 
   \frac{15}{2} \,u_3^{\text{(4)}} +\frac{33}{4} \,u_2^{\text{(5)}},    \\ 
  \dot u_5 & =&  -\,u_4  \,u_3' -\frac{1}{2} \,u_5  \,u_2' +\frac{1}{2} \,u_2  \,u_5' -\frac{3}{2} \,u_3  \,u_3'' +\,u_4  \,u_2'' +3 \text{u6}'' +\frac{7}{4}
   \,u_3  \,u_2''' -2 \,u_2  \,u_3''' +\,u_5''' +
       \notag \\  && 
   \frac{5}{2} \,u_2  \,u_2^{\text{(4)}} -3 \,u_3^{\text{(5)}} +4 \,u_2^{\text{(6)}},   \notag \\ 
  \dot u_6 & =&  \frac{1}{2} \,u_2  \text{u6}' -\frac{1}{2} \,u_5  \,u_3' -\frac{1}{2} \,u_4  \,u_3'' +\frac{3}{4} \,u_5  \,u_2'' -\frac{1}{2} \,u_3  \,u_3''' +\frac{3}{4}
   \,u_4  \,u_2''' +\text{u6}''' +\frac{3}{4} \,u_3  \,u_2^{\text{(4)}} -
       \notag \\  &&  \frac{1}{2} \,u_2  \,u_3^{\text{(4)}} +\frac{3}{4} \,u_2  \,u_2^{\text{(5)}} -\frac{1}{2} \,u_3^{\text{(6)}} +\frac{3}{4}
   \,u_2^{\text{(7)}}\notag.
 \eea
CS equations of motion at $z=3$: 
 \bea
 \dot \alpha_2 &= & -\frac{81}{10}  \,\alpha_2   \,\alpha_2' +\frac{3888}{35}  \,\alpha_4' -\frac{27}{20}  \,\alpha_2'''  , \notag \\ 
 \dot \alpha_3 &= & -\frac{405}{14}  \,\alpha_3   \,\alpha_2' -\frac{405}{14}  \,\alpha_2   \,\alpha_3' +\frac{540}{7}  \,\alpha_5' -\frac{27}{14}  \,\alpha_3'''  , \notag \\ 
 \dot \alpha_4 &= & \frac{59}{60}  \,\alpha_2'   \,\alpha_2'' +\frac{24}{5}  \,\alpha_2 {}^2  \,\alpha_2' -\frac{557}{30}  \,\alpha_2   \,\alpha_4' -\frac{152}{3}  \,\alpha_3  \,\alpha
   _3' -\frac{80}{3}  \,\alpha_4   \,\alpha_2' +\frac{100}{3}  \,\alpha_6' +
 \notag \\  &&  
   \frac{13}{30}  \,\alpha_2   \,\alpha_2''' -\frac{17}{30}  \,\alpha_4''' +\frac{1}{120}  \,\alpha_2^{\text{(5)}}  ,  \\ 
 \dot \alpha_5 &= & \frac{97}{140}  \,\alpha_3'   \,\alpha_2'' +\frac{29}{56}  \,\alpha_2'   \,\alpha_3'' +\frac{144}{35}  \,\alpha_2 {}^2  \,\alpha_3' +\frac{396}{35}  \,\alpha_3 
    \,\alpha_2   \,\alpha_2' -\frac{85}{14}  \,\alpha_2   \,\alpha_5' -\frac{252}{5}  \,\alpha_4   \,\alpha_3' -
         \notag \\  &&  
    \frac{1188}{35}  \,\alpha_3   \,\alpha_4' -
\frac{35}{2}  \,\alpha_5  \,\alpha
   _2' +\frac{5}{28}  \,\alpha_2   \,\alpha_3''' +\frac{123}{280}  \,\alpha_3   \,\alpha_2''' -\frac{1}{14}  \,\alpha_5''' +\frac{1}{560}  \,\alpha_3^{\text{(5)}}  , \notag \\ 
 \dot \alpha_6 &= & \frac{9}{25}  \,\alpha_3'   \,\alpha_3'' +\frac{79}{140}  \,\alpha_4'   \,\alpha_2'' +\frac{19}{56}  \,\alpha_2'   \,\alpha_4'' +\frac{80}{21} \,\alpha
   _2 {}^2  \,\alpha_4' +\frac{976}{105}  \,\alpha_3   \,\alpha_2   \,\alpha_3' +\frac{196}{15}  \,\alpha_4   \,\alpha_2   \,\alpha_2' +
     \notag \\  &&  
   \frac{55}{6}  \,\alpha_2   \,\alpha_6' +\frac{45}{7} \,\alpha
   _3 {}^2  \,\alpha_2' 
   -\frac{972}{35}  \,\alpha_4   \,\alpha_4' -\frac{224}{5}  \,\alpha_5   \,\alpha_3' -\frac{120}{7}  \,\alpha_3   \,\alpha_5' +\frac{41}{420}  \,\alpha_2  \,\alpha
   _4''' +
     \notag \\  &&  
   \frac{92}{525}  \,\alpha_3   \,\alpha_3''' +\frac{7}{15}  \,\alpha_4   \,\alpha_2''' +\frac{1}{6}  \,\alpha_6''' +\frac{ \,\alpha_4^{\text{(5)}} }{1680} \notag.
 \eea
KdV equations of motion at $z=4$:
\bea
 \dot u_2 & =&  \frac{8}{3} \,u_2' {}^2-\frac{4}{3} \,u_3  \,u_2' -\frac{4}{3} \,u_2  \,u_3' +4 \,u_5' +\frac{8}{3} \,u_2  \,u_2'' -4 \,u_4'' +\frac{2}{3}
   \,u_3''' +\frac{8}{3} \,u_2^{\text{(4)}},   \notag \\ 
  \dot u_3 & =&  \frac{40}{3} \,u_2'  \,u_2'' +\frac{8}{9} \,u_2 {}^2 \,u_2' -\frac{4}{3} \,u_2  \,u_4' -\frac{2}{3} \,u_2'  \,u_3' -\frac{4}{3} \,u_3 
   \,u_3' -\frac{4}{3} \,u_4  \,u_2' +4 \,u_6' -
       \notag \\  && 
   \frac{2}{3} \,u_2  \,u_3'' +6 \,u_5'' +\frac{16}{3} \,u_2  \,u_2''' -\frac{28}{3} \,u_4''' +\frac{13}{3}
   \,u_3^{\text{(4)}} +\frac{34}{9} \,u_2^{\text{(5)}},  \notag \\ 
  \dot u_4 & =&  \frac{40}{3} \,u_2'  \,u_2''' +\frac{4}{3} \,u_2  \,u_2' {}^2+\frac{2}{3} \,u_3  \,u_2  \,u_2' +\frac{4}{3} \,u_2  \,u_5' -\frac{4}{3} \,u_4 
   \,u_3' -\frac{2}{3} \,u_2'  \,u_4' -\frac{4}{3} \,u_3  \,u_4' -
       \notag \\  && 
   \frac{2}{3} \,u_5  \,u_2' +\frac{4}{3} \,u_2 {}^2 \,u_2'' -\frac{10}{3} \,u_2 
   \,u_4'' +10 \,u_2'' {}^2+\frac{2}{3} \,u_4  \,u_2'' +6 \,u_6'' +\frac{4}{3} \,u_2  \,u_3''' +\,u_3  \,u_2''' +
       \notag \\  && 
   4 \,u_5''' +\frac{14}{3} \,u_2 
   \,u_2^{\text{(4)}} -9 \,u_4^{\text{(4)}} +6 \,u_3^{\text{(5)}} +\frac{5}{3} \,u_2^{\text{(6)}},    \\ 
  \dot u_5 & =&  \frac{8}{3} \,u_2  \,u_2'  \,u_2'' +\frac{20}{3} \,u_2'  \,u_2^{\text{(4)}} +\frac{4}{9} \,u_4  \,u_2  \,u_2' +\frac{4}{3} \,u_2  \,u_6' +\frac{2}{3}
   \,u_3  \,u_2' {}^2-\frac{4}{3} \,u_4  \,u_4' -\frac{2}{3} \,u_5  \,u_3' -
       \notag \\  && 
   \frac{2}{3} \,u_2'  \,u_5' +\frac{2}{3} \,u_3  \,u_5' +\frac{40}{3}
   \,u_2''  \,u_2''' +\frac{2}{3} \,u_3  \,u_2  \,u_2'' +\frac{2}{3} \,u_2  \,u_5'' +\frac{2}{3} \,u_4  \,u_3'' -2 \,u_3  \,u_4'' +\frac{2}{3} \,u_5 
   \,u_2'' +
       \notag \\  && 
   \frac{8}{9} \,u_2 {}^2 \,u_2''' -\frac{8}{3} \,u_2  \,u_4''' +\frac{4}{3} \,u_3  \,u_3''' +\frac{4}{9} \,u_4  \,u_2''' +4 \,u_6''' +2 \,u_2 
   \,u_3^{\text{(4)}} +\frac{1}{3} \,u_3  \,u_2^{\text{(4)}} +\,u_5^{\text{(4)}} +
       \notag \\  && 
   \frac{14}{9} \,u_2  \,u_2^{\text{(5)}} -4 \,u_4^{\text{(5)}} +\frac{10}{3} \,u_3^{\text{(6)}},   \notag \\ 
  \dot u_6 & =&  \frac{2}{3} \,u_3  \,u_2'  \,u_2'' +\frac{8}{9} \,u_2  \,u_2'  \,u_2''' +\frac{4}{3} \,u_2'  \,u_2^{\text{(5)}} +\frac{2}{9} \,u_5  \,u_2 
   \,u_2' +\frac{2}{9} \,u_4  \,u_2' {}^2-\frac{2}{3} \,u_5  \,u_4' -
       \notag \\  && 
   \frac{2}{3} \,u_2'  \,u_6' +\frac{2}{3} \,u_3  \,u_6' +\frac{10}{3} \,u_2'' 
   \,u_2^{\text{(4)}} +\frac{2}{3} \,u_2  \,u_2'' {}^2+\frac{2}{9} \,u_4  \,u_2  \,u_2'' +\frac{2}{3} \,u_2  \,u_6'' -\frac{2}{3} \,u_4  \,u_4'' +
       \notag \\  && 
   \frac{2}{3} \,u_5 
   \,u_3'' +\frac{2}{9} \,u_3  \,u_2  \,u_2''' +\frac{20}{9} \,u_2''' {}^2+\frac{2}{3} \,u_4  \,u_3''' -\frac{2}{3} \,u_3  \,u_4''' -\frac{1}{9} \,u_5 
   \,u_2''' +\frac{2}{9} \,u_2 {}^2 \,u_2^{\text{(4)}} -\frac{2}{3} \,u_2  \,u_4^{\text{(4)}} +
       \notag \\  && 
   \frac{2}{3} \,u_3  \,u_3^{\text{(4)}} -\frac{1}{9} \,u_4  \,u_2^{\text{(4)}} +\,u_6^{\text{(4)}} +\frac{2}{3} \,u_2 
   \,u_3^{\text{(5)}} -\frac{1}{9} \,u_3  \,u_2^{\text{(5)}} +\frac{1}{9} \,u_2  \,u_2^{\text{(6)}} -\frac{2}{3} \,u_4^{\text{(6)}} +\frac{2}{3} \,u_3^{\text{(7)}} -\frac{1}{9} \,u_2^{\text{(8)}}\notag.
\eea

CS equations of motion at $z=4$:

\bea
 \dot \alpha_2 &= & \frac{2048}{21}  \,\alpha_3   \,\alpha_2' +\frac{2048}{21}  \,\alpha_2   \,\alpha_3' -\frac{4608}{7}  \,\alpha_5' +\frac{256}{21}  \,\alpha_3'''  , \notag \\ 
 \dot \alpha_3 &= & -\frac{160}{21}  \,\alpha_2'   \,\alpha_2'' -\frac{1280}{63}  \,\alpha_2 {}^2  \,\alpha_2' +\frac{1440}{7}  \,\alpha_2   \,\alpha_4' +\frac{5072}{21}  \,\alpha_3  \,\alpha
   _3' +\frac{1440}{7}  \,\alpha_4   \,\alpha_2' -
        \notag \\  &&  
   \frac{1800}{7}  \,\alpha_6' -\frac{80}{21}  \,\alpha_2   \,\alpha_2''' +\frac{90}{7}  \,\alpha_4''' -\frac{10}{63}  \,\alpha_2^{\text{(5)}} 
   , \notag \\ 
 \dot \alpha_4 &= & -\frac{272}{27}  \,\alpha_3'   \,\alpha_2'' -\frac{244}{27}  \,\alpha_2'   \,\alpha_3'' -\frac{128}{3}  \,\alpha_2 {}^2  \,\alpha_3' -\frac{832}{9}  \,\alpha_3 
    \,\alpha_2   \,\alpha_2' +\frac{1504}{27}  \,\alpha_2   \,\alpha_5' +\frac{896}{3}  \,\alpha_4   \,\alpha_3' +     \notag \\  &&  
    \frac{896}{3}  \,\alpha_3   \,\alpha_4' 
    +\frac{2800}{27}  \,\alpha_5  \,\alpha
   _2' -\frac{104}{27}  \,\alpha_2   \,\alpha_3''' -\frac{52}{9}  \,\alpha_3   \,\alpha_2''' +\frac{8}{9}  \,\alpha_5''' -\frac{2}{27}  \,\alpha_3^{\text{(5)}}  ,  \\ 
 \dot \alpha_5 &= & \frac{1108}{315}  \,\alpha_2   \,\alpha_2'   \,\alpha_2'' -\frac{884}{105}  \,\alpha_3'   \,\alpha_3'' -\frac{549}{140}  \,\alpha_4'  \,\alpha
   _2'' -\frac{51}{28}  \,\alpha_2'   \,\alpha_4'' +\frac{13}{168}  \,\alpha_2'   \,\alpha_2^{\text{(4)}} +\frac{256}{35}  \,\alpha_2 {}^3  \,\alpha_2' -
        \notag \\  &&  
   \frac{1504}{105}  \,\alpha_2 {}^2
    \,\alpha_4' -\frac{2720}{21}  \,\alpha_3   \,\alpha_2   \,\alpha_3' 
    -\frac{6208}{105}  \,\alpha_4   \,\alpha_2   \,\alpha_2' -\frac{800}{21}  \,\alpha_2   \,\alpha_6' +\frac{62}{63} \,\alpha
   _2' {}^3-
        \notag \\  &&  
   \frac{528}{7}  \,\alpha_3 {}^2  \,\alpha_2' +\frac{1944}{5}  \,\alpha_4   \,\alpha_4' +\frac{448}{3}  \,\alpha_5   \,\alpha_3' +\frac{1088}{21}  \,\alpha_3  \,\alpha
   _5' +
   \frac{47}{360}  \,\alpha_2''   \,\alpha_2''' +\frac{244}{315}  \,\alpha_2 {}^2  \,\alpha_2''' -
        \notag \\  &&  
   \frac{19}{42}  \,\alpha_2   \,\alpha_4''' -\frac{156}{35}  \,\alpha_3  \,\alpha
   _3''' -\frac{156}{35}  \,\alpha_4   \,\alpha_2''' -\frac{10}{7}  \,\alpha_6''' +\frac{29  \,\alpha_2   \,\alpha_2^{\text{(5)}} }{1260}
   -\frac{1}{280}  \,\alpha_4^{\text{(5)}} +\frac{ \,\alpha_2^{\text{(7)}} }{5040}
   , \notag \\ 
 \dot \alpha_6 &= & \frac{11828  \,\alpha_2   \,\alpha_3'   \,\alpha_2'' }{4725}+\frac{8902  \,\alpha_2   \,\alpha_2'   \,\alpha_3'' }{4725}+\frac{3673  \,\alpha_3   \,\alpha_2' 
    \,\alpha_2'' }{1050}-\frac{104}{25}  \,\alpha_4'   \,\alpha_3'' -\frac{56}{25}  \,\alpha_3'   \,\alpha_4'' +
         \notag \\  &&  
    \frac{32}{63}  \,\alpha_5'   \,\alpha_2'' +\frac{116}{63}
    \,\alpha_2'   \,\alpha_5'' 
    +\frac{451  \,\alpha_3'   \,\alpha_2^{\text{(4)}} }{9450}+\frac{41  \,\alpha_2'   \,\alpha_3^{\text{(4)}} }{1890}+\frac{128}{21}  \,\alpha_2 {}^3 \,\alpha
   _3' +\frac{2624}{105}  \,\alpha_3   \,\alpha_2 {}^2  \,\alpha_2' +
        \notag \\  &&  
   \frac{3200}{189}  \,\alpha_2 {}^2  \,\alpha_5' -\frac{2080}{21}  \,\alpha_4   \,\alpha_2   \,\alpha_3' 
   -\frac{1824}{35} \,\alpha
   _3   \,\alpha_2   \,\alpha_4' +\frac{448}{135}  \,\alpha_5   \,\alpha_2   \,\alpha_2' +\frac{6577  \,\alpha_2' {}^2  \,\alpha_3' }{3150}-
        \notag \\  &&  
   \frac{608}{7}  \,\alpha_3 {}^2 \,\alpha
   _3' -\frac{752}{7}  \,\alpha_3   \,\alpha_4   \,\alpha_2' +\frac{1728}{5}  \,\alpha_5   \,\alpha_4' 
   +\frac{1152}{7}  \,\alpha_4   \,\alpha_5' -\frac{1936}{21}  \,\alpha_3  \,\alpha
   _6' +
        \notag \\  &&  
   \frac{559  \,\alpha_3''   \,\alpha_2''' }{9450}+\frac{8}{175}  \,\alpha_2''   \,\alpha_3''' +\frac{1538  \,\alpha_2 {}^2  \,\alpha_3''' }{4725}+\frac{2459  \,\alpha_3  \,\alpha
   _2   \,\alpha_2''' }{1575}
   +\frac{152}{189}  \,\alpha_2   \,\alpha_5''' -
        \notag \\  &&  
   \frac{664}{175}  \,\alpha_4   \,\alpha_3''' -\frac{108}{175}  \,\alpha_3   \,\alpha_4''' -\frac{392}{135}  \,\alpha_5  \,\alpha
   _2''' +\frac{1}{189}  \,\alpha_2   \,\alpha_3^{\text{(5)}} +\frac{131  \,\alpha_3   \,\alpha_2^{\text{(5)}} }{6300}+\frac{2}{315}  \,\alpha_5^{\text{(5)}} +\frac{ \,\alpha_3^{\text{(7)}} }{37800} \notag.
 \eea


\bigskip
\noindent
\underline{\boldmath$N=7$}

\medskip
\noindent
CS-KdV map:
\bea
 u_2 &= & \,\, 56  \,\alpha_2  ,  \notag \\ 
 u_3 &=  & \,\, 140  \,\alpha_2' -504  \,\alpha_3  ,  \notag \\ 
 u_4 &=  & \,\, -1008  \,\alpha_3' +168  \,\alpha_2'' +784  \,\alpha_2 {}^2+4320  \,\alpha_4  ,  \notag \\ 
 u_5 &=  & \,\, 2352  \,\alpha_2   \,\alpha_2' +6480  \,\alpha_4' -864  \,\alpha_3'' +112  \,\alpha_2''' -8928  \,\alpha_2   \,\alpha_3 -31680  \,\alpha_5  ,   \\ 
 u_6 &=  & \,\, -8928  \,\alpha_2   \,\alpha_3' +1180  \,\alpha_2' {}^2-8928  \,\alpha_3   \,\alpha_2' -31680  \,\alpha_5' +1408  \,\alpha_2   \,\alpha_2'' +3600  \,\alpha_4'' -
        \notag \\  &&  
 360
    \,\alpha_3''' 
    +40  \,\alpha_2^{\text{(4)}} +2304  \,\alpha_2 {}^3+40320  \,\alpha_4   \,\alpha_2 +18000  \,\alpha_3 {}^2+172800  \,\alpha_6  ,  \notag \\ 
 u_7 &=  & \,\, 708  \,\alpha_2'   \,\alpha_2'' +3456  \,\alpha_2 {}^2  \,\alpha_2' +20160  \,\alpha_2   \,\alpha_4' -4488  \,\alpha_2'   \,\alpha_3' +18000  \,\alpha_3  \,\alpha
   _3' +
          \notag \\  &&  
   20160  \,\alpha_4   \,\alpha_2' +86400  \,\alpha_6' 
   -2544  \,\alpha_2   \,\alpha_3'' -2664  \,\alpha_3   \,\alpha_2'' -8640  \,\alpha_5'' +312  \,\alpha_2  \,\alpha
   _2''' +
          \notag \\  &&  
   720  \,\alpha_4''' -60  \,\alpha_3^{\text{(4)}} +6  \,\alpha_2^{\text{(5)}} -13824  \,\alpha_3   \,\alpha_2 {}^2
   -103680  \,\alpha_5   \,\alpha_2 -86400  \,\alpha_3   \,\alpha_4 -
          \notag \\  &&  
518400  \,\alpha_7  \notag.
 \eea
KdV equations of motion at $z=2$:
\bea
  \dot u_2 & =&  2 \,u_3' -5 \,u_2'',   \notag \\ 
  \dot u_3 & =&  -\frac{10}{7} \,u_2  \,u_2' +2 \,u_4' +\,u_3'' -10 \,u_2''',   \notag \\ 
  \dot u_4 & =&  -\frac{8}{7} \,u_3  \,u_2' +2 \,u_5' -\frac{20}{7} \,u_2  \,u_2'' +\,u_4'' -10 \,u_2^{\text{(4)}},   \notag \\ 
  \dot u_5 & =&  -\frac{6}{7} \,u_4  \,u_2' +2 \,u_6' -\frac{12}{7} \,u_3  \,u_2'' +\,u_5'' -\frac{20}{7} \,u_2  \,u_2''' -6 \,u_2^{\text{(5)}},   \\ 
  \dot u_6 & =&  -\frac{4}{7} \,u_5  \,u_2' +2 \,u_7' -\frac{6}{7} \,u_4  \,u_2'' +\,u_6'' -\frac{8}{7} \,u_3  \,u_2''' -\frac{10}{7} \,u_2  \,u_2^{\text{(4)}} -2
   \,u_2^{\text{(6)}} ,  \notag \\ 
  \dot u_7 & =&  -\frac{2}{7} \,u_6  \,u_2' -\frac{2}{7} \,u_5  \,u_2'' +\,u_7'' -\frac{2}{7} \,u_4  \,u_2''' -\frac{2}{7} \,u_3  \,u_2^{\text{(4)}} -\frac{2}{7} \,u_2 
   \,u_2^{\text{(5)}} -\frac{2}{7} \,u_2^{\text{(7)}} \notag.
    \eea
CS equations of motion at $z=2$:
\bea
 \dot \alpha_2 &= & -18  \,\alpha_3'  , \notag \\ 
 \dot \alpha_3 &= & \frac{8}{3}  \,\alpha_2   \,\alpha_2' -\frac{120}{7}  \,\alpha_4' +\frac{1}{6}  \,\alpha_2'''  , \notag \\ 
 \dot \alpha_4 &= & \frac{10}{3}  \,\alpha_3   \,\alpha_2' +\frac{12}{5}  \,\alpha_2   \,\alpha_3' -\frac{44}{3}  \,\alpha_5' +\frac{1}{15}  \,\alpha_3'''  , \notag \\ 
 \dot \alpha_5 &= & \frac{14}{5}  \,\alpha_3   \,\alpha_3' +4  \,\alpha_4   \,\alpha_2' +\frac{16}{7}  \,\alpha_2   \,\alpha_4' -\frac{120}{11}  \,\alpha_6' +\frac{1}{28} \,\alpha
   _4'''  ,  \\ 
 \dot \alpha_6 &= & \frac{16}{5}  \,\alpha_4   \,\alpha_3' +\frac{18}{7}  \,\alpha_3   \,\alpha_4' +\frac{14}{3}  \,\alpha_5   \,\alpha_2' +\frac{20}{9}  \,\alpha_2   \,\alpha_5' -6 \,\alpha
   _7' +\frac{1}{45}  \,\alpha_5'''  , \notag \\ 
 \dot \alpha_7 &= & \frac{20}{7}  \,\alpha_4   \,\alpha_4' +\frac{18}{5}  \,\alpha_5   \,\alpha_3' +\frac{22}{9}  \,\alpha_3   \,\alpha_5' +\frac{16}{3}  \,\alpha_6   \,\alpha_2' +\frac{24}{11}
    \,\alpha_2   \,\alpha_6' +\frac{1}{66}  \,\alpha_6''' \notag  .
 \eea
KdV equations of motion at $z=3$:
\bea
  \dot u_2 & =&  -\frac{12}{7} \,u_2  \,u_2' +3 \,u_4' -6 \,u_3'' +4 \,u_2''',  \notag \\ 
  \dot u_3 & =&  -\frac{12}{7} \,u_3  \,u_2' -\frac{12}{7} \,u_2  \,u_3' +3 \,u_5' +3 \,u_4'' -14 \,u_3''' +15 \,u_2^{\text{(4)}},   \notag \\ 
  \dot u_4 & =&  -\frac{12}{7} \,u_3  \,u_3' -\frac{9}{7} \,u_4  \,u_2' +\frac{3}{7} \,u_2  \,u_4' +3 \,u_6' +\frac{6}{7} \,u_3  \,u_2'' -\frac{30}{7} \,u_2 
   \,u_3'' +3 \,u_5'' +
     \notag \\  && 
   \frac{30}{7} \,u_2  \,u_2''' +\,u_4''' -15 \,u_3^{\text{(4)}} +21 \,u_2^{\text{(5)}} ,  \notag \\ 
  \dot u_5 & =&  -\frac{9}{7} \,u_4  \,u_3' -\frac{6}{7} \,u_5  \,u_2' +\frac{3}{7} \,u_2  \,u_5' +3 \,u_7' -\frac{18}{7} \,u_3  \,u_3'' +\frac{9}{7} \,u_4 
   \,u_2'' +3 \,u_6'' +\frac{24}{7} \,u_3  \,u_2''' -
     \notag \\  && 
   \frac{30}{7} \,u_2  \,u_3''' +\,u_5''' +\frac{45}{7} \,u_2  \,u_2^{\text{(4)}} -9 \,u_3^{\text{(5)}} +15 \,u_2^{\text{(6)}} ,
     \\ 
  \dot u_6 & =&  -\frac{6}{7} \,u_5  \,u_3' -\frac{3}{7} \,u_6  \,u_2' +\frac{3}{7} \,u_2  \,u_6' -\frac{9}{7} \,u_4  \,u_3'' +\frac{9}{7} \,u_5  \,u_2'' +3
   \,u_7'' -\frac{12}{7} \,u_3  \,u_3''' +\frac{15}{7} \,u_4  \,u_2''' +
     \notag \\  && 
   \,u_6''' +3 \,u_3  \,u_2^{\text{(4)}} -\frac{15}{7} \,u_2  \,u_3^{\text{(4)}} +\frac{27}{7} \,u_2 
   \,u_2^{\text{(5)}} -3 \,u_3^{\text{(6)}} +\frac{39}{7} \,u_2^{\text{(7)}} ,  \notag \\ 
  \dot u_7 & =&  -\frac{3}{7} \,u_6  \,u_3' +\frac{3}{7} \,u_2  \,u_7' -\frac{3}{7} \,u_5  \,u_3'' +\frac{6}{7} \,u_6  \,u_2'' -\frac{3}{7} \,u_4  \,u_3''' +\frac{6}{7} \,u_5 
   \,u_2''' +\,u_7''' -\frac{3}{7} \,u_3  \,u_3^{\text{(4)}} +\frac{6}{7} \,u_4  \,u_2^{\text{(4)}} +
     \notag \\  && 
   \frac{6}{7} \,u_3  \,u_2^{\text{(5)}} -\frac{3}{7} \,u_2  \,u_3^{\text{(5)}} +\frac{6}{7} \,u_2 
   \,u_2^{\text{(6)}} -\frac{3}{7} \,u_3^{\text{(7)}} +\frac{6}{7} \,u_2^{\text{(8)}} \notag .
   \eea
CS equations of motion at $z=3$:
\bea
\dot \alpha _2{}   & =&  -12 \,\alpha _2   \,\alpha _2'   +\frac{1620}{7} \,\alpha _4'   -2 \,\alpha _2''' ,   \notag \\ 
\dot  \alpha _3{}   & =&  -\frac{300}{7} \,\alpha _3   \,\alpha _2'   -\frac{300}{7} \,\alpha _2   \,\alpha
   _3'   +\frac{1320}{7} \,\alpha _5'   -\frac{20}{7} \,\alpha _3'''  ,  \notag \\  
\dot \alpha _4{}   & =&  \frac{24}{5}
   \,\alpha _2'    \,\alpha _2  {}^2-32 \,\alpha _4'    \,\alpha _2  +\frac{13}{30} \,\alpha _2'''   \,\alpha
   _2  -44 \,\alpha _4   \,\alpha _2'   -\frac{379}{5} \,\alpha _3   \,\alpha _3'   +120 \,\alpha
   _6'   +\frac{59}{60} \,\alpha _2'    \,\alpha _2''  -  \notag \\ && 
   \,\alpha _4'''  +\frac{1}{120} \,\alpha
   _2{}^{(5)},  \notag \\  
\dot \alpha _5{}   & =&  \frac{144}{35} \,\alpha _3'    \,\alpha _2  {}^2+\frac{396}{35} \,\alpha _3  
   \,\alpha _2'    \,\alpha _2  -\frac{1488}{77} \,\alpha _5'    \,\alpha _2  +\frac{5}{28} \,\alpha _3'''  
   \,\alpha _2  -\frac{420}{11} \,\alpha _5   \,\alpha _2'   -\frac{882}{11} \,\alpha _4   \,\alpha
   _3'   -
    \notag \\ &&
   \frac{4392}{77} \,\alpha _3   \,\alpha _4'   +\frac{540}{11} \,\alpha _7'   +\frac{97}{140} \,\alpha
   _3'    \,\alpha _2''  +\frac{29}{56} \,\alpha _2'    \,\alpha _3''  +\frac{123}{280} \,\alpha _3  
   \,\alpha _2'''  -\frac{25}{77} \,\alpha _5'''  +\frac{1}{560} \,\alpha _3{}^{(5)}   ,  \\ 
\dot \alpha _6{}   & =& 
   \frac{80}{21} \,\alpha _4'    \,\alpha _2  {}^2+\frac{196}{15} \,\alpha _4   \,\alpha _2'    \,\alpha
   _2  +\frac{976}{105} \,\alpha _3   \,\alpha _3'    \,\alpha _2  -4 \,\alpha _6'    \,\alpha _2  +\frac{41}{420}
   \,\alpha _4'''   \,\alpha _2  +\frac{45}{7} \,\alpha _3  {}^2 \,\alpha _2'   -
    \notag \\ &&
   24 \,\alpha _6   \,\alpha
   _2'   -\frac{396}{5} \,\alpha _5   \,\alpha _3'   -54 \,\alpha _4   \,\alpha _4'   -\frac{275}{7} \,\alpha
   _3   \,\alpha _5'   +\frac{79}{140} \,\alpha _4'    \,\alpha _2''  +\frac{9}{25} \,\alpha _3'   
   \,\alpha _3''  +\frac{19}{56} \,\alpha _2'    \,\alpha _4''  +
    \notag \\ &&
   \frac{7}{15} \,\alpha _4   \,\alpha
   _2'''  +\frac{92}{525} \,\alpha _3   \,\alpha _3'''  +\frac{\,\alpha _4{}^{(5)}  }{1680},  \notag \\  
\dot \alpha_7{}   & =&  \frac{40}{11} \,\alpha _5'    \,\alpha _2  {}^2+\frac{816}{55} \,\alpha _5   \,\alpha _2'   
   \,\alpha _2  +\frac{3996}{385} \,\alpha _4   \,\alpha _3'    \,\alpha _2  +\frac{1940}{231} \,\alpha _3   \,\alpha
   _4'    \,\alpha _2  +\frac{156}{11} \,\alpha _7'    \,\alpha _2  +\frac{61}{990} \,\alpha _5'''   \,\alpha
   _2  +
    \notag \\ &&
   \frac{304}{21} \,\alpha _3   \,\alpha _4   \,\alpha _2'   +\frac{77}{15} \,\alpha _3  {}^2 \,\alpha _3'   -72
   \,\alpha _6   \,\alpha _3'   -\frac{324}{7} \,\alpha _5   \,\alpha _4'   -\frac{220}{7} \,\alpha _4   \,\alpha
   _5'   -20 \,\alpha _3   \,\alpha _6'   +
    \notag \\ &&
   \frac{65}{132} \,\alpha _5'    \,\alpha _2''  +\frac{443
   \,\alpha _4'    \,\alpha _3''  }{1540}+\frac{103}{440} \,\alpha _3'    \,\alpha _4''  +\frac{491 \,\alpha
   _2'    \,\alpha _5''  }{1980}+\frac{83}{165} \,\alpha _5   \,\alpha _2'''  +\frac{69}{385} \,\alpha _4  
   \,\alpha _3'''  +
    \notag \\ &&
   \frac{25}{264} \,\alpha _3   \,\alpha _4'''  +\frac{2}{11} \,\alpha _7'''  +\frac{\,\alpha
   _5{}^{(5)}  }{3960}.  \notag 
 \eea

\end{document}